# Hierarchical materials with interconnected pores from capillary suspensions for bone tissue engineering

Souhaila Nider* [1], Femke De Ceulaer [1], Berfu Göksel [2], Annabel Braem [2], Erin Koos* [1]

[1] KU Leuven, Department of Chemical Engineering, 3001 Leuven, Belgium

[2] KU Leuven, Department of Materials Engineering, 3001 Leuven, Belgium

* Corresponding authors: erin.koos@kuleuven.be, souhaila.nider@kuleuven.be

**Abstract:** The increasing demand for bone grafts due to the aging population has opened new opportunities for the manufacture of porous ceramics to assist in bone reconstruction. In our study, we investigate a new, promising method to manufacture hierarchically porous structures in a straightforward, and tuneable way. It consists of combining the novel technology of capillary suspensions, formed by mixing solid particles and two immiscible liquids, one less than 5 vol%, with freeze casting. We have successfully achieved alumina and $\beta$-TCP materials with both <2 µm and 20-50 µm as the smallest and largest pore size, respectively. The microstructure exhibits fully open pores and high levels of porosity (> 60%). The capillary suspensions' rheological behaviour indicates that silica nano-suspensions as a secondary fluid creates a stronger internal particle network than sucrose for the alumina system. Conversely, the opposite was observed with the $\beta$-TCP system. These differences were attributed to the change in affinity between the secondary fluids and the solid loading. Our study, both systems has served to deepen the knowledge about the new area of capillary suspensions and proves their use in hierarchical porous scaffolds for bone tissue engineering.

**Graphical abstract:**

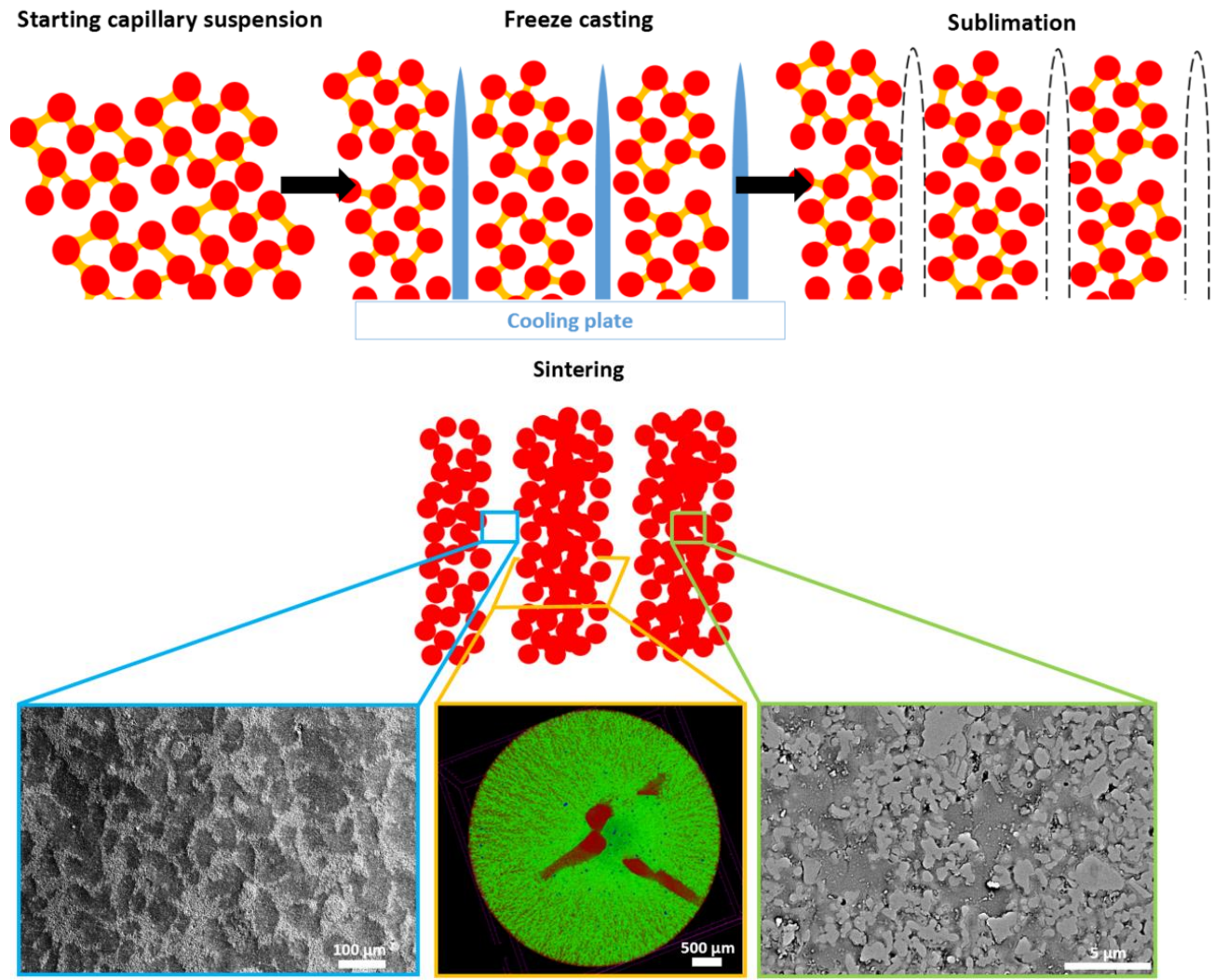

# 1 Introduction

While bone possesses inherent regenerative capabilities, severe fractures and extensive damage may hinder natural healing and lead to further complications, particularly in older individuals with limited regenerative potential [1]. About 2 million bone graft surgeries per year are needed worldwide, underlying the critical need to developing performant implants [2,3,4,5]. As a result, significant attention has been directed toward the exploration of implant matrices sourced from synthetic ceramic materials [6,7,1,8,9,10].

Alumina is largely present in orthopaedic and dental applications due to its good mechanical strength and excellent corrosion resistance in the body physiological environment [11][12]. It inherently possesses low osseointegration and osteogenic capacities, which can, nonetheless, be overcome by adding porosity or a bioactive coating, as demonstrated by Camilo et al. [11]. Despite their outstanding role in medical devices [13][14], further research is necessary to optimise the design and overcome the low fracture toughness in porous alumina to ensure good bone anchorage and avoid implant instability and failure [14][15].

Beta-tricalcium phosphate (β-TCP) has gained an increasing amount of interest over the years due to its superior degradation rate following insertion [16][17][18][19] as well as its strong chemical resemblance with the inorganic part of the bone [6][9][16]. While β-TCP possesses inferior mechanical properties compared to alumina, it is not inert. In fact, numerous studies have shown that β-TCP promotes osteoinduction[9][16][20][17][21][22][18]. Clinically, pure β-TCP is typically used in maxillofacial surgery [10][23], where it has been proven to form new bone with uninflamed connective tissue and blood vessels, without any necrosis nor foreign body reaction noted [23]. One of the main challenges with β-TCP is its high dissolution rate, which can be mitigated through an adequate design [10][19][23]. Putri et al. [24] studied the bone regeneration process after the implantation of a dual porous β-TCP scaffold in a rabbit's femoral region. They demonstrated that β-TCP can successfully be used as bone grafts but they also emphasized the importance of having a hierarchical pore size for bone regeneration.

Various methodologies are available for fabricating materials with hierarchical architectures, mainly classified into soft templating, hard templating, and template-free methods [25,26,27]. Soft templating utilizes self-assembled molecular structures such as micelles and vesicles to guide pore formation [28,29,30,31,32,33]. Surfactants link these aggregates, enabling the creation of hierarchical structures with precisely controlled pore sizes. Emulsion templating, a variation where emulsions serve as templates, offers an additional level of pore size control [25,34]. However, these methods require complex chemical reactions and precise control over emulsion stability, which reveals being challenging. Moreover, the obtained pore sizes and interconnection are insufficient for bone tissue engineering. Hard templating employs sacrificial materials or replicas to define pores [35,36,37,38,39,40,41]. Sacrificial templating offers precise control over pore size and shape but may introduce safety risks and limit pore interconnectivity. Replica methods can tailor the pore structure but are susceptible to crack formation during processing [42,43]. Finally, template-free methods involve techniques like partial sintering and 3D printing. While these offer versatility, they may result in lower porosity and require careful control of processing parameters and ink properties such as its viscosity, storage and loss moduli, etc. [35,36,44,45,46,47,48,49].

The technique to build porous materials highlighted in this work is freeze casting [50,51,52,53]. This method, notable for its environmental friendliness and simplicity, involves the freezing of a suspension through the application of a, typically unidirectional, thermal gradient. The solvent undergoes nucleation which leads to the formation of dendrites or lamellae, depending on the freezing agent. They grow along the thermal gradient, progressively displacing the suspended particles between them until the entire suspension is fully frozen. After freeze drying to sublime the solvent, the green body is left with voids previously occupied by the freezing agent and retains its structural integrity due to the binders initially present within the suspension. Finally, the structure is consolidated through (partial) sintering. Freeze casting is a promising technique investigated in bone tissue engineering [54,55,56,57,58,59]. The resulting scaffolds possess a high porosity, interconnectivity and directionality, which has been proven to improve both *in vitro* cell proliferation and *in vivo* bone regeneration when compared to non-freeze-casted scaffolds [60][54]. The final pore size and porosity are determined not only by the starting powder dimensions but also by the thermal gradient used, offering great potential for pore control [56][55][61][62] and adaptability to the patients. Unfortunately, linking the initial suspension's rheology to the final scaffold morphology [62], achieving precise control over multimodal pore size distributions [63] and addressing the decrease in scaffold strength with increasing porosity [53][55][56] are significant challenges in producing porous ceramics via freeze casting. In an effort to address the challenges associated with the fabrication of porous ceramics and achieve enhanced control over pore size distribution, we propose the integration of capillary suspensions.

Capillary suspensions represent a relatively recent area of research, constituting a three-phase system comprised of two immiscible liquids and a solid phase [64][65][66][67]. One of the liquids, the secondary liquid, is present in a limited quantity, typically accounting for less than 5% of the suspension volume. Depending on the wetting characteristics of the secondary and bulk liquids with respect to the solid particles, two distinct states can be defined: the pendular state and the capillary state. In the pendular state, the three-phase contact angle between the secondary fluid and the solid inside the bulk liquid is less than 90°. This implies that the secondary fluid exhibits superior wetting properties towards the solid compared to the bulk fluid. Consequently, the secondary fluid forms concave bridges between individual particles. As the volume of secondary fluid increases, the system becomes saturated. This transition to a funicular state, where the bridges span multiple particles, is characterized by a plateau in the shear moduli. Conversely, the capillary state arises when the three-phase contact angle exceeds 90°. This results in a microstructure characterised by small secondary liquid droplets surrounded by multiple particles [68]. Regardless of whether the system is in the capillary or the pendular state, the resultant microstructure comprises a sample-spanning network of particles made by the secondary fluid, suspended in the bulk fluid [69]. An example of both states are presented in supplemental Figure S1 [70][71].

Capillary suspensions have been previously used to create porous ceramics from alumina by Weiss [72], highlighting the versatility offered by the capillary suspension route in the control of the porosity, strength and pore size of the resulting material. As a consequence, they constitute a potential solution to the limitations of freeze casting, namely the difficulties in achieving defined and controlled multimodal pore size distributions and by maintaining porous body strength with increasing porosity. Specifically, they can address these drawbacks by introducing additional porosity while maintaining strength.

While literature already exists about the freeze casting of polymeric solutions and suspensions [54,55,56,57,58,73], our study focusses on the feasibility of integrating capillary suspensions with freeze casting to fabricate hierarchically porous materials. It has the advantage of being a promising method to manufacture them in an eco-friendly, straightforward, tunable and easy way for bone tissue engineering. Combining these methods offers the potential to produce personalised bone grafts by altering many parameters influencing the porosity and pore size such as the thermal gradient, particle nature, secondary fluid, etc. To our knowledge, this is the first time that these two technologies are combined. Capillary suspensions yield to a porous microstructure [74] after partial sintering, while freeze casting introduces additional, directional pores within this framework [52]. Alumina is tested first as the solid loading and serves as a model material for a more biocompatible powder: β-TCP. Several configurations of secondary fluids and powders with different particle size are investigated and their influence on both the rheological properties of the paste and the resulting pore morphology are discussed.

## 2 Materials and methods

### 2.1 Materials

Commercially available alumina powder with an average particle size of 0.5 μm (CT3000SG, Almatis GmbH, Germany) and 5 μm β-TCP agglomerated microspheres (BABI-TCP-SP, Berkeley Advanced

|  | Composition | Density (g/ml) | $d_{50}$ (nm) |
|---|---|---|---|
| Bulk fluid | Camphene | 0.85 | - |
| Solid | $Al_2O_3$ | 3.9 | 500 |
|  | β-TCP | 3.14 | 5000 |
| Secondary fluid | Sucrose solution 50 vol%: |  |  |
|  | Solid sucrose | 1.59 | - |
|  | mQ water | 1 |  |
|  | Self-made silica nano-suspension 10 vol%: |  |  |
|  | Aerosil 130 silica | 2.65 | 16 |
|  | mQ water | 1 | - |
|  | Commercial LUDOX TM-50 suspension | 1.4 | 12-13 |

*Table 1: Materials used to form capillary suspensions in this study.*

Biomaterials, USA) were chosen as the materials of interest. The raw starting materials were characterised by scanning electron microscopy (SEM, XL30, The Netherlands) and can be found in Figure S2. The alumina particles show an arbitrary isometric shape while the TCP granules exhibit approximately 1 μm pores. Three different secondary liquids were examined: a 50 wt% (approximately 30 vol%) aqueous suspension of amorphous silica in water (Ludox TM-50, $d_{50} = 12 - 13$ nm, Sigma-Aldrich, Germany), a self-prepared aqueous suspension containing 10 vol% of fused amorphous silica (Aerosil 130, $d_{50} = 16$ nm, Sigma Aldrich, Germany) dispersed through speed mixing for 5 min at 3500 rpm, and an aqueous preparation of 50 vol% sucrose (Sigma Aldrich, USA, Germany). The Ludox particles are electrostatically stabilised and freely flow whereas the Aerosil particles in the self-made suspension are unstabilised and form a gel (see Figure S3). Camphene (C10H16, Sigma-Aldrich, Belgium), used as received, was the freezing agent. The material properties for each component of the capillary suspension preparation are displayed in Table 1.

## 2.2 Capillary suspension preparation

There are several methods for preparing a capillary suspension. Combining the particles with the bulk fluid using a high-shear dissolver mixer before adding the secondary liquid is the most prevalent approach for preparing capillary suspensions [65 68 71 75 76]. Conversely, the mixing order can be reversed by first combining the two fluids [77]. Factors such as mixing speed, mixing order, and quantities significantly affect the suspension quality. The most effective method for our system is detailed below. Further explanations for selecting this specific method, which differs from the common preparation techniques, are provided in the Supplemental Information.

The capillary suspensions were prepared at 60°C. This temperature was selected due to camphene's melting point being 48°C, and to the established procedure of preparing camphene-based suspensions for freeze casting at 60°C in literature [56, 78, 79]. Neither alumina nor TCP powders undergo any phase transitions at this temperature. Prior to mixing, the powders were heated in an oven and the liquids warmed together in the same container inside a water bath. Ten minutes after the temperature reached 60°C, the fluids were emulsified by using an ultrasonic mixer (UP400S, Hielscher, Germany) at 20% of its maximum power (400W). Two secondary liquid types were used in this study: a sucrose solution, and silica nano-suspension suspensions. The secondary liquid serves two key functions. First, it creates a particle network by either directly bridging the particles (sucrose) or creating small particle clusters that are in turn networked together (the silica nano-suspension). Second, it must stabilise the green body during debinding and sintering. The sucrose can either crystallize or form a glass whereas the nanoparticles in the suspension should link the microparticles via van der Waals attraction. Both will result in mechanical stability during debinding and sintering. Since these two secondary fluids have completely different natures — one a solution and the other a colloidal suspension — we anticipate that their effects on the final scaffold will be significantly different. Based on the work of Weiss et al.[80], the silica-based secondary liquid should increase the porosity and enhance mechanical strength beyond that of the sucrose-based secondary liquid. As for the sucrose, it eventually gels, creating an even more cohesive network.. During the (partial) sintering process, even if the sucrose is eliminated by the temperature, the particles remain close, resulting in a connected network in the dry state.

Immediately after the mixing of the two fluids, the pre-heated particles (7 or 10 vol% solid loading) were added. The powder temperature was verified prior to adding it to the emulsion since the camphene could otherwise recrystallise around the particles, hindering the formation of capillary bridges. The evaporation of water at the increased temperature is negligible due to the small miscibility of water in camphene.

After the three components were combined, they were returned to the water bath where they were mixed for approximately 3 h at 700 rpm using a propeller mixer to form a homogeneous sample. After mixing, the suspensions were poured into a stainless steel mould (25 mm diameter and 80 mm height). The room temperature (20.5 $\pm$ 1 °C) mould was closed at the bottom with aluminium foil. The samples were open to the atmosphere on the top.

The scaffolds were demoulded after 1 h and subjected to a 24-h sublimation process at room temperature under local ventilation. The green bodies were partially sintered at a maximum temperature of 1400 °C for alumina and 1075 °C for β-TCP. Those low maximum temperatures were intentionally chosen to avoid complete sintering. Indeed, while these low temperatures do contribute to the future secondary porosity, the secondary liquids used in the capillary suspensions also play a significant role. By acting as bridges between the particles, they influence both the degree of porosity and the pore size. As for $\beta$-TCP, we additionally wanted to avoid its transition to $\alpha$-TCP, occurring at 1125°C, since this phase is more soluble when introduced in the body environment. Therefore, we chose a lower temperature of 1075°C to account for any potential error in the furnace temperature.

The sintering procedure involved a heating rate of 5 °C/min up to 500 °C, followed by a 10 °C/min ramp to 800 °C, and finally reaching 1400 °C for alumina or 1075 °C for β-TCP at a rate of 5 °C/min. This maximum temperature was maintained for 3 h. The cooling process to room temperature followed the inverse sequence: from the maximum temperatures (1400°C for alumina and 1075°C for TCP), the temperature was decreased to 800°C at a rate of 5°C/min, then further reduced to 500°C at a rate of 10°C/min, and finally cooled to room temperature at a rate of 5°C/min. We selected this cooling procedure to prevent crack formation due to rapid cooling in our samples. The second step of the heating rate was increased in the 500-800°C range purely because of the furnace operational limitations. The sintering occurred under ambient atmosphere due to the anticipated lack of oxygen's specific impact on both alumina and TCP.

## 2.3 Freeze casting

Camphene, possessing a crystallisation temperature of 48°C, was used as the primary fluid and freezing vehicle without further purification. The freeze casting experiments conducted in this study were performed at room temperature ($20.5 \pm 1$ °C). After preparation, the capillary suspension was poured in a stainless steel mould (25 mm diameter and 80 mm height), resulting in the formation of a controlled pore size. Freezing at room temperature and without further temperature control was chosen for both ease and as it produced pores of approximately 100 µm, as shown in the preliminary experiments where multiple cooling rates were investigated (Figure S4).

## 2.4 Characterisation

The β-TCP powder phase was ascertained using x-ray crystallography. Measurements were taken within a $2\theta$ range of 2-80°, utilising a step size of 0.02° and a running time of 20.7 minutes. Supplemental Figure S5 depicts the results between 25 and 35°. The analysis indicated the presence of 61 vol% β-TCP together with hydroxyapatite (18 vol%) and other additional calcium phosphate phases. The fitting of the patterns was realised with $\beta$-TCP (COD ID 1529466 and 9005865) and hydroxyapatite (2300273). More detail about the percentage of the other phases and the fitting are shown in the supplementary information. Thermal gravimetric analysis (TGA, Q500, TA Instrument, USA) was additionally performed on the raw powder (Figure S6). This measurement shows that either no organic (polymeric) coating is present on the β-TCP particles or that this coating is negligibly thin since the weight decrease in the 20-500°C range was within the error limits of the device.

Microstructure analysis was carried out on infiltrated samples. Following the partial sintering treatment, the densified scaffolds were infiltrated with epoxy resin (EpoFix kit, Struers, Germany) and subjected to degassing. The degassing process comprised three successive vacuum cycles at approximately 10 mbar, where the final cycle lasted for 1 h. After undergoing a 12 h epoxy curing process at room temperature, the approximatively 20 mm high and 25 mm diameter samples were meticulously sectioned transversally in their middle using a diamond saw to reveal their internal microstructure. Multiple polishing steps were executed subsequently to achieve a smoother surface. The obtained sections were first manually polished using P80, P180, and P360 grade sandpaper, then underwent more refined polishing utilising a semi-automatic polishing machine (Struers, Germany) with consecutively 15, 3, and 1 µm diamond suspensions.

The microstructure was characterised using scanning electron microscopy (SEM, Jeol, Japan, and SEM, XL30, The Netherlands). To enhance the samples conductivity, a thin layer of Au/Pd was applied on their surface (20 s at 40 mA). Certain scaffolds were subjected to additional analysis through X-ray micro-tomography (TeScan Unitom XL, Czech Republic) at a voltage of 150 kV, a power output of 15 W and an exposure duration of 450 ms. A 1 mm aluminium plate was positioned in front of the detector to serve as an electron beam filter, and 3000 projections were acquired per 360° rotation. The voxel size for all samples analysed ranged from 3 to 3.7 µm, as measured via the device's AcquilaTM software.

Image analysis was conducted on the obtained SEM pictures using both Python and the open access software ImageJ. Porosity analysis was executed using a Python code (see supplementary information) where an Otsu filter was applied to the images to determine the ratio between black and white pixels, representing voids and solid loading walls, respectively. More precise pore characteristics, like the pore size, the Feret diameter and the Feret angle, were determined after image processing, noise removal and adjustment of the brightness using ImageJ. Mercury infiltration porosimetry (MIP) (AutoPore, Micromeritics instrument corporation, USA) was used to determine the internal pore size and porosity of the samples, with a 0.5 to $3 \times 10^4$ PSIA pressure range.

Prior to freeze casting, the rheological properties of the capillary suspension pastes were tested using an ARG2 rheometer from TA Instruments (USA). A Couette Peltier setup was used with a vane geometry to measure the properties at a temperature of 53°C. Oscillatory amplitude stress sweeps with amplitudes ranging from 0.01 to 10% at a frequency of 1 Hz and frequency sweeps from 100 to 0.1 rad/s were carried out. The frequency sweeps used a strain amplitude of 0.1%, verified to be within the linear viscoelastic region (LVE) for all samples. To minimize camphene evaporation, a solvent trap was used for all rheological measurements.

The paste was additionally characterised using a confocal microscope (Leica TCS SP8, 63x glycerol objective, Germany) to investigate the arrangement between the particles and the secondary liquid inside the main fluid. To highlight the secondary liquid, rhodamine B isothiocyanate (RBITC, Sigma-Aldrich, Germany) was added to the secondary liquid during preparation and excited using a 552 nm laser. The three-phase contact angle of the alumina and β-TCP systems at 60°C was measured using a sessile drop setup (Biolin Scientific, Attension Theta Flex, Sweden). A droplet of the secondary liquid was deposited on either an alumina or β-TCP (obtained by conventional pressing and sintering at 1100 °C for 10 h at a heating rate of 10 °C/min) plate inside a camphene environment and measured using OneAttension software, provided by Biolin Scientific. We define the contact angle as the angle between the horizontal solid surface below the secondary liquid sessile drop and the tangent to the drop's curvature at the contact point, as shown in Figure S7a.

## 3    Results and discussion

### 3.1    Three-phase contact angle

The three-phase contact angle is a crucial parameter for determining the state of capillary suspensions, whether pendular or capillary, as it influences the microstructure of the suspension [70,71]. The distinction between the pendular and capillary states is essential due to their significant impact on the suspension rheology and final scaffold's morphological characteristics. Contact angle measurements on the alumina plate reveal that the sucrose three-phase contact angle is 72 ± 6°, which is notably higher than the Ludox silica nano-suspension (50 ± 6°), as shown in Figure S7 a-c. The self-made silica nano-suspension had a contact angle of 54 ± 3°. The three systems possess a contact angle lower than 90°, placing them in the pendular state, with the particle network in the alumina-sucrose system expected to be weaker than the alumina-silica system, given its higher contact angle.

The β-TCP system is found to be in the capillary state with angles measuring 137 ± 4° for the sucrose, 116 ± 2° for the Ludox and 126 ± 12° for the self-made nano-suspension system (Figure S7 d-f). Given the significant difference in β-TCP affinity between the sucrose and the silica fluids, a stronger capillary network in the silica case is anticipated. Since the alumina and the β-TCP systems lead to a different state (pendular for the alumina and capillary for the β-TCP), we expect significant differences in their rheological behaviour and microstructure.

It is important to disclose that when the three-phase contact angle experiments involving the self-made silica nano-suspension were realised, it was noticed that the droplet gelled during the time span of a few minutes. Supplemental Figure S3 was taken after physically moving the droplets that were initially on the alumina plate inside a camphene environment.

## 3.2 Capillary suspension pastes

The integration of freeze casting with capillary suspensions to realise hierarchical porous materials is investigated first with alumina powder before moving to β-TCP to examine the extension of such technology for bioactive materials.

### 3.2.1 Alumina-based capillary suspensions

To demonstrate the existence of a capillary suspension in our experiments, a sample of 7% alumina and 1% of rhodamine-B dyed sucrose was imaged using confocal microscopy. Figure 1 presents the incorporation of dyed sucrose, depicted in red, acting as the secondary fluid positioned between the alumina particles. The undyed particles are shown as shadows in Figure 1b, some of which are highlighted using blue circles. In the lower magnification image (Figure 1a), the sucrose is uniformly dispersed throughout different areas of the suspension, indicating homogeneity within the capillary suspension. Distinct secondary liquid droplets measuring approximately 0.2 – 1 µm appear to connect alumina particles (indicated by the blue circles in Figure 1b). The bridged particles should form a sample-spanning network within the sample, modifying the mechanical properties of the paste.

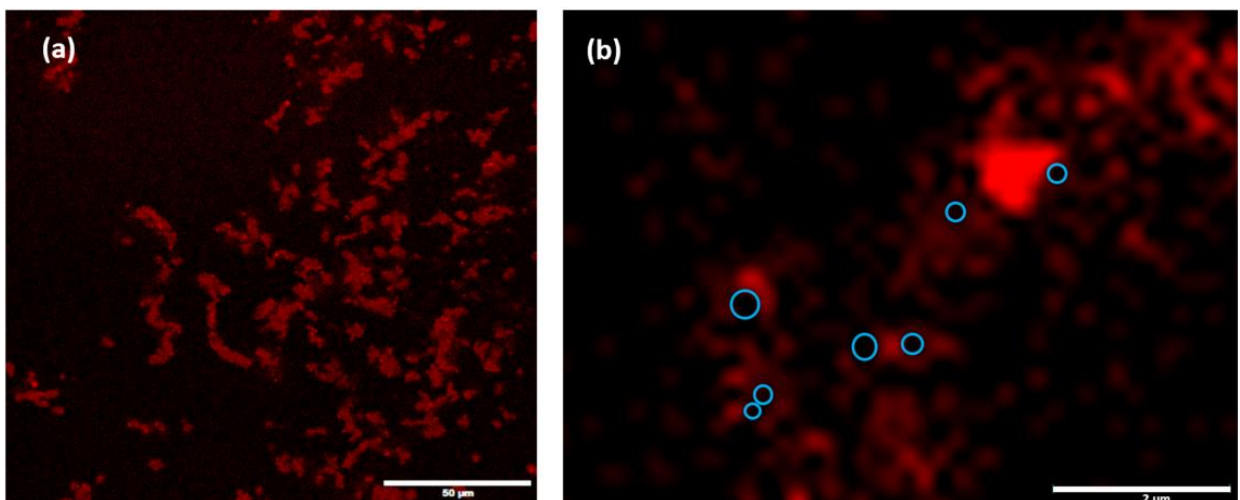

*Figure 1: Confocal microscopy image of a capillary suspension containing 7 vol% of undyed alumina and 1 vol% aqueous sucrose dyed with rhodamine (appearing in red). (a) The sucrose is homogeneously distributed inside the suspension where it forms (b) distinct drops, bridging the alumina particles. Some of these particles are highlighted using blue circles.*

Capillary suspensions with diverse secondary fluid natures and concentrations are rheologically characterised to establish a correlation between their flow properties and composition. Shear oscillatory amplitude sweeps are used to assess the presence of a sample-spanning particle network via an increase in elasticity, as previously shown for capillary suspensions by Koos et al.[64,66,68,70]. By applying a gradual oscillatory stress, the magnitude of the shear modulus and critical strain marking the end of the linear viscoelastic region (LVE), both of which are directly linked to the internal structure of the suspension, can be determined. The LVE defines the range within tests can be conducted without disrupting the material's structure, that is the yield stress for the material, indicating either a more flexible or rigid network. The storage modulus represents the elastic component of a viscoelastic material, reflecting energy stored by the network, while the loss modulus accounts for the dissipated energy of the network and surrounding fluid. A higher storage modulus is often indicative of a robust internal network. In the case of capillary suspensions, the increase of secondary liquid provokes an increase in the moduli values in the LVE [64,71,77]. When secondary liquid is added, the number of bridges joining particles inside the suspension increases, strengthening the structure, which can be seen in the rheological data by an increase in the moduli and by network formation in the microscopy analysis (Figure 1). However, this increase in moduli with the secondary fluid concentration is only valid up until a critical concentration denoting the transition to the funicular state where the secondary liquid saturates the particles without any further bridge formation, leading to a weakening in both the network and moduli. After the LVE, non-linearity with the applied stress begins to play a significant role [81,82] but this should not be relevant to the processes used here.

The amplitude sweep graph is depicted in Figure 2a ($G'$: storage modulus) and 2b ($G''$: loss modulus) for an alumina-based capillary suspension (solid loading of $\phi = 7$ vol%) with a self-made silica nano-suspension (in green) and a sucrose solution (in blue) as secondary fluids. As a reference point, the rheological behaviour of the pure suspension was also examined (in red), giving a storage modulus in the LVE of 1200 ± 300 Pa and a loss modulus of 600 ± 100 Pa. It is noteworthy that the highest values for both the storage and loss moduli, 2400 ± 300 Pa and 1100 ± 300 Pa for 1 or 2 vol%, respectively, are found for the silica nano-suspension as secondary fluid. On the other hand, the use of sucrose led to a slight decrease in moduli in comparison to the pure suspension, with a storage and loss modulus of 500 ± 300 Pa and 300 ± 100 Pa, respectively.

In this study, the formation of silica nano-suspension bridges between the alumina particles seems to engender a stronger network compared to the one resulting from its substitution with sucrose, as evidenced by their lower moduli. Despite being slight, a discernible trend can be observed in the critical stress at the end of the LVE between the sucrose and the silica nano-suspension curves: the yield stress is the highest when the nano-suspension is used as the secondary fluid, followed by the pure suspension, and lowest when sucrose is used. The yield stress of a material is defined as the stress at which it starts to flow, and it is directly related to the internal structure of the material. Stronger inter-particle attractive forces result in a higher yield stress, allowing the material to resist deformation more effectively. This resistance to deformation is nicely captured under oscillatory stress conditions where both the inter-particle strength and its dependence on the oscillatory amplitude is captured. In essence, a higher yield stress indicates better green body stability, as it prevents the particles from easily separating or settling and ensures better preservation of moulded or printed forms.

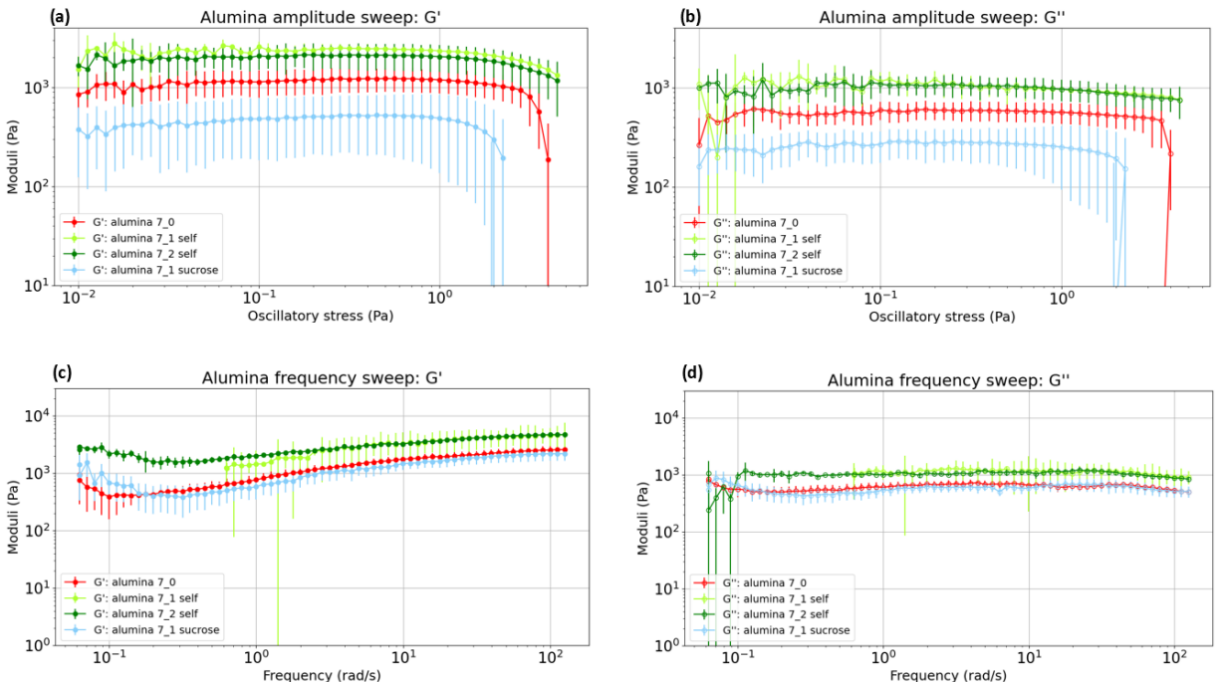

*Figure 2: Rheological characterisation of 7 vol% alumina solid loading capillary suspensions with different concentrations of secondary liquids (a) amplitude sweep – $G'$, (b) amplitude sweep – $G''$, (c) frequency sweep – $G'$ (d) frequency sweep – $G''$. The notation "self" stands for "self-made silica nano-suspension".*

This implies that the most stable suspension is obtained with the nano-suspension of silica, whereas the weakest is the sucrose capillary suspension. This is in line with the three-phase contact angle measurements, where it was demonstrated that the angle changed from 54 ± 3° for the silica nano-suspension, to 72 ± 6° for the sucrose, signifying a higher affinity of the silica nano-suspensions to the alumina compared to the sucrose. As the reason why the sucrose-based capillary suspension has even lower moduli than the pure suspension (without secondary fluid added), it seems that the network obtained with the sucrose is easily broken while the compact structure of the pure suspension exerts more stress on the vane. Even if the breaking and reformation of bridges should increase the moduli, it is not the case with the sucrose in our experiments. In consequence, this would imply few and weak interactions of the alumina particles with the sucrose.

When the 1 vol% and 2 vol% self-made silica nano-suspensions are compared, their curves do not exhibit any significant differences in the moduli, suggesting that either the network strength is similar or that rheological measurements are not precise enough to capture the differences in their micro-structure. In capillary suspension rheology, even if the amount of secondary liquid differs, the moduli can reach the same value if the capillary suspension lies in the funicular state [64,65,66]. The shear modulus is a measure of both the inter-particle force (capillary bridge strength) and the number of connections (network structure). Initially, with the addition of small amounts of secondary liquid, pendular bridges form between particle pairs. The small bridge size leads to a weak connection, a weak internal network, resulting in a lower yield stress. As the secondary fluid concentration increases, the number and size of these bridges increases, strengthening the suspension's yield stress and moduli. This continues until a critical point where the suspension transitions from the pendular state to the funicular state. The funicular state is characterised by secondary fluid bridges spanning multiple particles (e.g. trimers). Because the microstructure remains similar at the end of the pendular state and the beginning of the funicular state and the capillary force to displace any single particle in the cluster plateaus, the shear modulus remains unchanged, indicating the transition between these two states. As shown in Figure 2a, the shear modulus for the 1 and 2 vol% self-made suspension as secondary fluid are nearly identical, a sign of the pendular to funicular transition. As for the differences in the microstructure, this will be further investigated in section 3.3.

Furthermore, a frequency sweep was performed on the same suspensions, as depicted in Figure 2c-d. The storage modulus exhibits a slight increase with frequency for all curves, which is characteristic of a weak gel [83][84], confirming the presence of a particle network within the capillary suspensions. The slight increase of the storage modulus with frequency indicates a slight dependence of the aggregate formation and particle packing with shear rate (as the angular frequency is related to the maximum transient shear rate). These aggregates offer greater resistance to flow, leading to an increase in the storage modulus. The loss modulus (Figure 2d) remains frequency independent, indicating stability in energy dissipation with frequency. From a quantitative perspective, the suspensions containing silica exhibited higher moduli, confirming that the strength of the particle network is the greatest among the tested suspensions.

Overall, the rheological data demonstrate the superior performance of the silica-based capillary suspensions, highlighting their stronger particle network and higher stability. This is likely due to the affinity between the alumina and the different secondary liquids: the silica nano-suspension droplets exhibited a lower three-phase contact angle than the sucrose, indicating a stronger affinity between the silica suspensions and the alumina. Since the alumina particles are better wetted by the silica nano-suspensions, the capillary forces present in the alumina-silica system are stronger than the ones in the alumina-sucrose one, resulting in a stronger and more stable capillary suspension.

### 3.2.2 β–TCP-based capillary suspensions

The β-TCP-based capillary suspensions have also been rheologically characterised to determine the influence of the secondary fluid type and fraction. The resulting amplitude sweep curves are presented in Figure 3a (storage modulus) and 3b (loss modulus) showing the influence of 0, 0.5, 1, 1.5, and 2 vol% of aqueous sucrose, as well as 1 and 2 vol% of a self-made silica nano-suspension. Since the incorporation of secondary liquid into capillary suspensions should either form or strengthen the particle network, the moduli of capillary suspensions are expected to surpass those of the corresponding pure suspensions. This phenomenon is observed when using the sucrose as the secondary liquid, where the storage modulus increases from $2.7 \pm 0.9 \times 10^3$ Pa for the pure suspension to $1.0 \pm 0.1 \times 10^4$ Pa when aqueous sucrose is added. Since our system lies in the capillary state, it is suggested that the aqueous sucrose droplets in our system form convex bridges[70] between β-TCP particles, thereby establishing a network that resists collapse under shear. However, a decrease in the moduli occurs for the silica nano-suspension compared to the pure suspension. This implies that the tested concentrations of 1 and 2 vol% have either no or a detrimental impact on the network, meaning that a weak network is formed when the self-made nano-suspension is added.

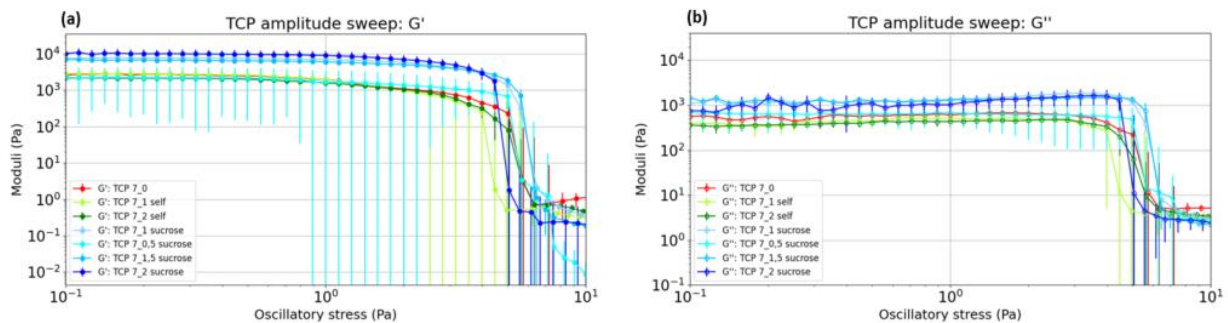

*Figure 3: β-TCP amplitude sweep showing (a) $G'$ and (b) $G''$ as a function of applied stress amplitude for 7% solid loading. The notation "self" stands for "self-made silica nano-suspension".*

The end of the LVE, which marks the yield stress of the material, varies depending on the secondary liquid used. The yield stress is approximately 1 Pa for the sucrose solution as a secondary fluid, while it only reaches 0.3 Pa for the pure and the silica suspensions. This suggests that not only is the particle network of sucrose-based capillary suspensions stronger, but it is also more stable than the other suspensions tested for this specific β-TCP system. Within the different concentrations of the same secondary liquid, the end of the LVE remains relatively constant.

Furthermore, the sucrose-based capillary suspensions (blue curves) exhibit a steep decline in their storage modulus at a stress of approximately 5-7 Pa. This drop is characteristic of suspensions with a yield stress and rapid network breakdown. In contrast, the silica-based capillary suspensions (green curves) and the pure suspension (red curve) display a more gradual decrease in their storage modulus, which rapidly decreases at around 3 Pa of applied stress. This larger yielding region indicates that the particle network breaks more gradually, likely due to a larger distribution in interaction forces. For the silica-based capillary suspension, this is probably caused by small bridges breaking, aligning the suspensions in the direction of flow and subsequently breaking larger bridges. For the pure suspension, this indicates the presence of a weak network that progressively breaks. This dependence on aggregation is confirmed via the loss modulus with the sucrose-based β-TCP suspension exhibiting a different behaviour than for other secondary fluids, as shown in Figure 3b. The sucrose-based system exhibits a gradual increase in $G''$ with stress amplitude before experiencing a sharp decline when the yield stress is reached. The increase in the loss modulus corresponds to increased friction between particles and can be linked to aggregate formation at elevated shear, opposing the shear flow, thereby increasing the loss modulus. Such behaviour is not observed in the other suspensions, indicating that their structures break down without aggregate formation. The frequency sweep (Figure S8) exhibits a similar trend as obtained with for the alumina suspensions indicating a weak gel structure.

## 3.3 Structure of sintered samples

Following the rheological characterisation of the capillary suspensions in their liquid state, our focus shifted to analysing the microstructure of the samples after freeze casting and sintering. We aimed to establish possible correlations between the rheological observations and the internal structure of the produced scaffolds.

### 3.3.1 Alumina porous bodies

The structure captured via x-ray micro-tomography of a sample with 7 vol% alumina with 1 vol% of self-made silica nano-suspension is shown in Figure 4. In the acquired images, voids representing porosity are depicted in red, while alumina particles with intermediate density range are displayed in green, and those with the highest density range, e.g. compact clusters, are rendered blue. Figure 4a illustrates an axial section, while Figure 4b presents a corresponding transverse slice. A schema at the bottom of each picture represents how the initial scaffold was sectioned. A directional pore network can be observed at the periphery of the sample with elongated pores extending from the surface towards the centre, while the core of the cylinder appears denser. The large pores present throughout are thought to be attributed to large air bubbles that were released during the solidification of the camphene. The elongated pores are attributed to the formation of dendrites characterised by long, branched lamellae resembling a tree-like structures (Figure S4). Camphene freezes in the direction of the thermal gradient; the capillary suspension at 60°C is poured into a room-temperature mould, immediately crystallizing the camphene at the edges, leading to radially-oriented dendritic structures. As the dendrites grow, they displace the solid particles, as was previously demonstrated by Yoon et al. [56], resulting in densification towards the centre. Moreover, the propagating dendrites exert a pressure on pre-existing air bubbles, leading to their coalescence and the formation of the observed large voids in Figure 4.

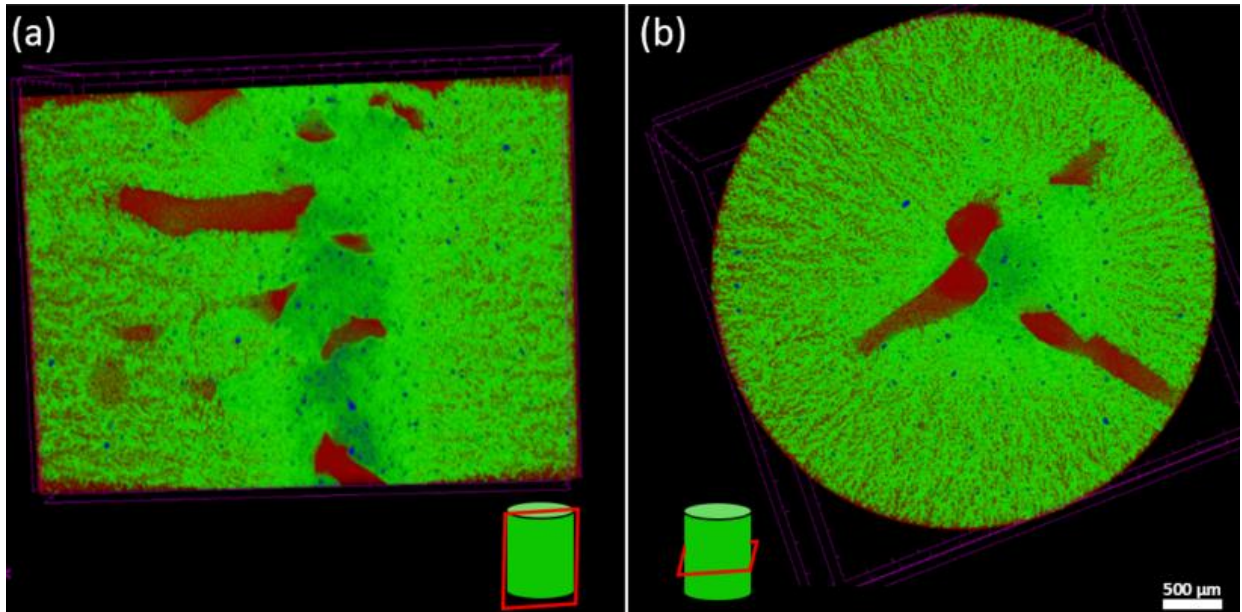

*Figure 4: X-ray micro computed tomography images of a 7 vol% alumina and 1 vol% of self-made silica nano-suspension processed by freeze casting and sintering. (a) Axial and (b) transverse cross section of the resulting alumina scaffold. Pores are indicated in red, solid material in green and high-density regions are in blue. A schema at the bottom of each picture represents how the initial scaffold was sectioned. The scale bar (500 µm) in (b) is the same for both images.*

To obtain a more detailed view of the microstructure, scanning electron microscopy (SEM) analysis is conducted, as shown in Figure 5, for bodies with same alumina fraction (7 vol%) but different volume fractions of the self-made silica nano-suspension as secondary fluid. In these images, the lighter regions represent alumina, while the darker areas correspond to voids (filled with epoxy resin). In the pure suspension (Figure 5a), the dendritic structures are clearly visible. They consist of primary lamellae with secondary branches. When 0.5 vol% and 1 vol% of secondary liquid are added (Figure 5b and 5c, respectively), minimal differences are observed, and the structures exhibit the same tree-like structure. However, the secondary branches are no longer present for the 2 vol% secondary fluid sample. Instead, the pores are ellipsoidal without clear branching. Interestingly, when the concentration of the secondary fluid is further increased to 5 vol%, the initial structure is re-established, such that dendrites with primary and secondary branches are visible. For all the concentrations, the presence of camphene induced interconnectivity between the pores and open porosity of the scaffolds. Freeze casting is a technique that inherently gives pores interconnectivity [53] [51] [55] [88] due to camphene dendrite formation. In our investigation, we primarily analysed transverse cuts, with additional longitudinal cuts also examined. Both types of cuts revealed apparent two-dimensional pore interconnectivity, suggestive of three-dimensional pore interconnectivity (as is shown in the x-ray tomography images of Figure 4). The epoxy-infiltrated samples measured approximately 20 mm in height with a 25 mm diameter are halved at their midpoint, revealing fully open and interconnected pores as evidenced by the complete epoxy infiltration within all the scaffolds.

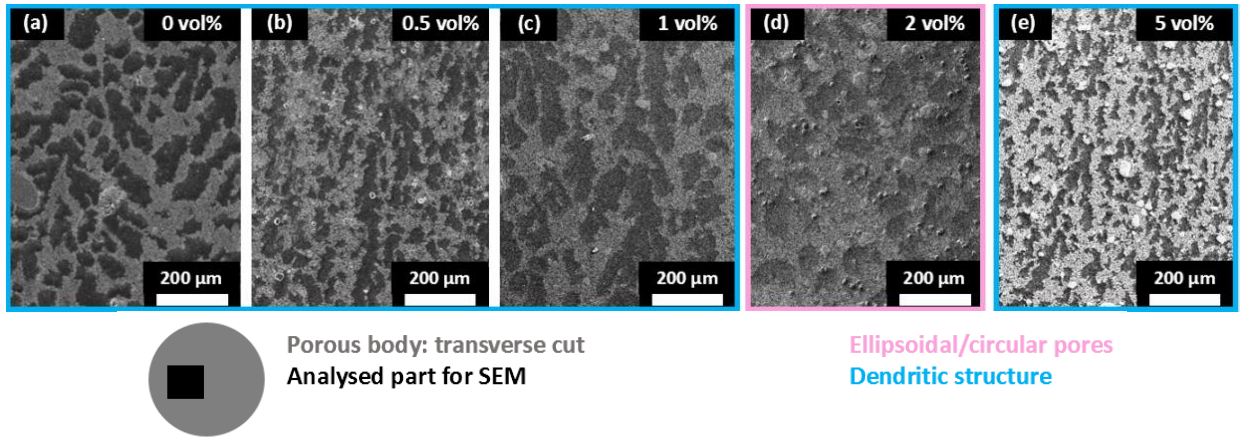

*Figure 5: SEM transverse cross sections of 7 vol% alumina scaffolds with (a) 0, (b) 0.5, (c) 1, (d) 2 and (e) 5 vol% of the self-prepared silica nano-suspension used as a secondary fluid. The cuts were taken in the transition zone between the centre and the edges of each scaffold. The brighter colour represents the alumina while the darker one corresponds to the voids. All the images were taken at the same magnification.*

Upon camphene solidification, the dendrites readily displace the alumina particles for secondary fluid concentrations lower than 2 vol%, giving rise to the primary and secondary dendrites in those samples. Thus, the capillary suspension pastes at concentrations below 2 vol% secondary fluid are too weak to withstand the force exerted by the camphene. Despite the plateau in network strength in the rheological data, the supposedly funicular bridges in the 2 vol% sample appear to hinder particle motion during camphene crystallisation. Consequently, the transformation into dendrites becomes more difficult for the camphene. As a result, the microstructure exhibits more circular and elongated pores.

Capillary suspension rheology for secondary fluids that preferentially wet the particles, e.g. samples in the pendular state, as is the case in for the alumina samples, typically shows an increase in the strength (yield stress or shear moduli) with small amounts of added liquid until the bridges become saturated, indicating a transition to the funicular state [64]. A schematic representation of the link between the microstructure obtained and the evolution of the network strength and yield stress with the secondary fluid is represented in Figure S9. In this study, we go further by examining the morphology after sintering. When the concentration of the secondary fluid is further increased to 5 vol%, the SEM pictures show that the initial dendritic structure re-emerges. This can be attributed to the alumina particles becoming saturated by the secondary liquid, which again weakens the structure. As a result, flocculation occurs within the suspension. This facilitates the formation of the characteristic branch-like structures by camphene with the tight spherical aggregates, as seen in Figure 5e.

To further investigate the influence of the secondary fluid on the resulting structures of the porous bodies, a sucrose solution was employed as an alternative to the self-made silica nano-suspension, as shown in Figure 6. As with the self-made silica, a transition from dendritic (Figure 6a and 6b) to ellipsoidal pores (Figure 6c) was observed when 2 vol% of sucrose was added, which then implies the presence of enhanced capillary forces between alumina particles resulting in a stronger particle network.

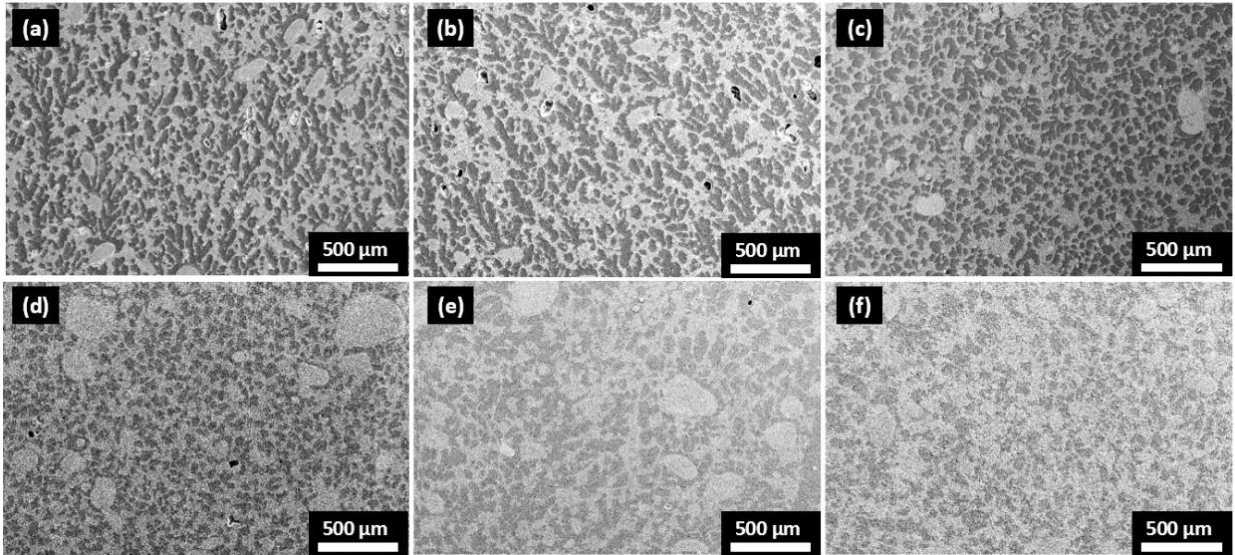

*Figure 6: SEM transverse cross section of 7 vol% alumina scaffolds with (a) 0, (b) 1 and (c) 2 vol% of sucrose solution used as a secondary fluid. SEM transverse cross section of 10 vol% alumina scaffolds with (d) 0, (e) 1 and (f) 2 vol% of sucrose solution used as a secondary fluid. The brighter colour represents the alumina while the darker area corresponds to the voids.*

Surprisingly, the rheological findings for the paste do not completely align with the sintered structures observed under scanning electron microscopy. Indeed, the capillary suspension with 2 vol% silica nano-suspension exhibited an apparently stronger network than the sucrose solution under rheology but seem to possess similar microstructures under SEM. Moreover, contrary to our expectations, the capillary suspension containing both 1 and 2 vol% of the self-made silica nano-suspension demonstrated the highest strength according to rheological measurements. These discrepancies with the microscopy analysis, where only 2 vol% seemed to give a stronger particle network, suggest that rheology may not possess sufficient precision to discern the distinctions between various concentrations in this context. However, rheology still proves that the capillary suspension using sucrose as a secondary liquid exhibited the weakest structure, even lower than the suspension without added secondary liquid, which is consistent with the three-phase contact angle measurements. One area where the network moduli may influence the sintered body is in the fragility of the sintered material. During their fabrication, delicate handling was required due to the fragility of the 1 vol% samples.

We additionally imaged fracture surfaces on select samples to corroborate the observations made with the infiltrated samples. These are presented in Supplementary Figure S10 for 7 vol% alumina samples with 0, 1, and 2 vol% sucrose, as well as 1 and 2 vol% self-made silica nano-suspension. In the binary system (pure suspension), the analysis revealed the characteristic dendritic morphology of the camphene surrounded by alumina particles, which leads to elongated pores. The addition of sucrose resulted in the formation of significantly larger and partially porous alumina aggregates, as can be observed in Figure S10b. This is likely due to sucrose promoting the agglomeration of individual alumina grains, but with insufficient secondary fluid to bridge these aggregates. Consequently, the microstructure resembled that of the pure suspension with larger particles. When the sucrose concentration was further increased (Figure S10c and S10f), the alumina aggregates remained but they appeared bridged by thinner and more porous alumina threads. This is hypothesised to occur in a two-step process: first, sucrose induces the formation of aggregates, and then, excess sucrose facilitates the connection between the remaining alumina particles, forming a more robust network. This network hinders the development of camphene branches, resulting in the observed rounded pores at 2 vol% sucrose, unlike the structures observed at lower secondary liquid concentrations. The distinction between primary and secondary porosity becomes the most evident at 2 vol% sucrose.

The incorporation of the self-made silica nano-suspension (Figure S10g-j) resulted in the formation of distinct microstructures, regardless of the secondary liquid concentration. Both 1 and 5 vol% exhibited agglomerates bridged by porous alumina. However, the 1 vol% suspension displayed larger and fewer agglomerates compared to the 5 vol% counterpart. This observation is attributed to the presence of a greater amount of secondary phase in the 5 vol% mixture, which likely facilitated a more homogeneous distribution during mixing, consequently leading to the formation of more numerous, but smaller, agglomerates.

To enhance their overall strength while maintaining an optimal level of porosity for bone tissue engineering, the alumina concentration was increased from 7 to 10 vol%. Figures 6d-6f display SEM images corresponding to 10 vol% alumina scaffolds with different concentrations of sucrose, namely 0, 1, and 2 vol% respectively. Notably, the pure suspension exhibits a change in the pore structure compared to the previous 7 vol% alumina case. In this instance, the pores appear round, without dendritic structures even for the pure suspension. However, the notable distinction between structures with and without the secondary fluid lies in the pore size distribution (as shown in Section 3.3.4).

Previously, we hypothesised that the presence of ellipsoidal pores formed during the freezing of camphene is indicative of a stronger network. In the case of the 7 vol% alumina, the strength of the network was attributed to an increase in the number of capillary bridges. However, in this scenario with 10 vol%, the increased strength (due to the absence of dendritic structures) can be attributed to the higher concentration of alumina, which increases the number of particle contacts. When a higher solid loading is utilised, the development of dendrites with secondary branches is hindered, suppressing the expected dendritic shape. The thermal gradient and growth of secondary branches is further tempered due to the increase in the thermal conductivity from the increased solid loading.

The microstructure obtained with our 10 vol% system closely resembles the one from Domenech and Velankar [89]. Despite using a different system (silica bridged with polyethylene oxide (PEO) dispersed in polyisobutylene), they observed visible pendular particle network with PEO menisci linking silica particle pairs. They additionally noticed large agglomerates in some regions of the suspension. They explained the formation of both the pendular network and the aggregates through the secondary fluid droplet size distribution during capillary suspension preparation, a hypothesis supported by Bossler et al. [90] investigating the influence of mixing conditions on pendular state suspensions. When a particle encounters a significantly larger drop during mixing, it gets engulfed due to the wettability, forming large agglomerates as this process repeats with other particles. Conversely, smaller droplets form layers around the particles, and subsequent encounters with other particles create concave bridges, resulting in a pendular structure. This may explain the aggregates present in our sucrose system.

As shown in the rheological measurements (Figure 2), the capillary suspensions with self-made suspensions as the secondary fluid exhibited higher shear moduli than both the pure suspension and suspension with sucrose as the secondary liquid. To investigate the influence of the particles and their stability in the secondary liquid, we compared the structure of 10 vol% alumina-based scaffolds obtained with a 2 vol% self-made silica nano-suspension as the secondary fluid with one obtained from a commercial silica nano-suspension. The primary distinction between these two silica nano-suspensions lies in their composition: the self-prepared nano-suspension comprises 10 vol% of 16 nm fumed silica particles, whereas the stabilised Ludox nano-suspension consists of 50 wt% (corresponding to approximately 30 vol%) of 12-13 nm silica particles. As shown in supplemental Figure S3, the self-made nano-suspension forms a gel-like structure at 10 vol% whereas the Ludox nano-suspension is still freely flowing at 50 wt% (Figure S7).

Upon the addition of the silica nano-suspensions to the 10 vol% alumina in camphene, a notable increase in the formation of aggregates is observed in both silica-based capillary nano-suspension samples (Figure 7) compared to the use of sucrose as a secondary fluid (Figure 6d-f).

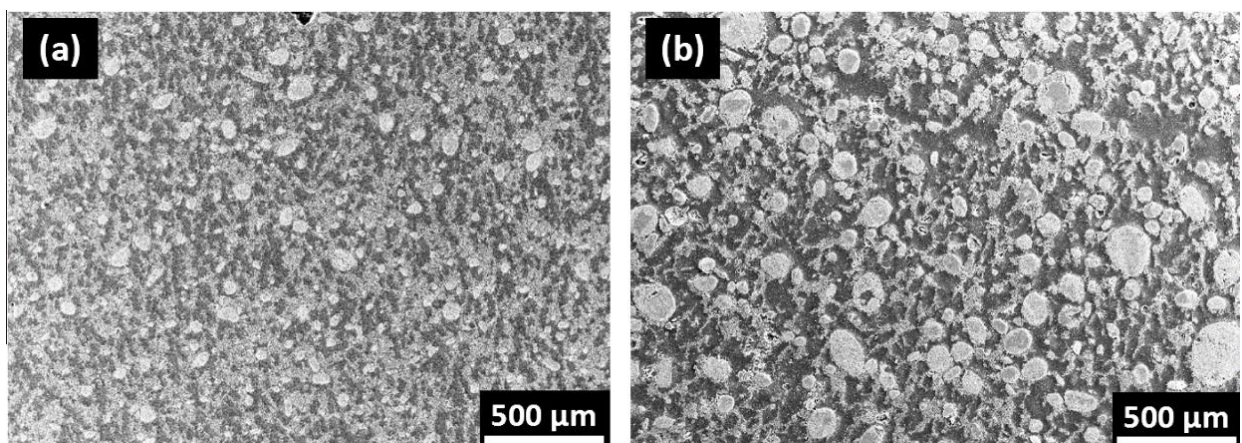

*Figure 7: SEM cross sections of 10 vol% alumina scaffolds with 2 vol% of (a) a self-prepared and (b) commercial silica nano-suspension as secondary fluid. The brighter colour represents the alumina while the darker one corresponds to the voids.*

Agglomerates are present in both the sucrose and silica systems, however, their morphologies differ significantly. The agglomerates in the silica system are more compact compared to the more porous ones in the sucrose system. We believe this difference arises from the lower contact angle produced by the silica nano-suspension on alumina. This lower contact angle causes the particles to come closer together, filling the gaps and forming denser aggregates. During the partial sintering of the wet capillary suspensions, silica remains between the alumina particles, physically bridging them and creating dense aggregates. This effect is enhanced by the localisation of the nanoparticles, supposed to not only be present in the pendular bridge, but also be distributed around individual alumina particles, as was demonstrated by Liu et. al. [86]. In their work, they showed confocal microscopy images of a capillary suspension composed of silica microparticles as the solid loading and a nano-suspension of silica as the secondary fluid. They proved that a part of the nanoparticles are localised around the microparticles. Their findings might contribute to both the lower moduli for the TCP and the denser aggregates formed by the alumina system in the sintering stage because of them being surrounded by silica nanoparticles. Moreover, Fusco et al. [87] demonstrated that when silica nanoparticles droplets are under motion, their spreading behaviour increased, which is the case during the mixing of our capillary suspensions and can be the cause of the presence of nanoparticles around alumina. Conversely, sucrose is eliminated by heat, which will result in more porous aggregates.

In both Figure 7a and b, the presence of large aggregates is evident, with these agglomerates being more pronounced when the Ludox nano-suspension is utilised. The reason behind this change remains unclear. Since both silica nano-suspensions possess a similar three-phase contact angle with alumina, a possible explanation may be that after the evaporation of water, the Ludox suspension contains more silica particles (because of higher initial silica content). Those can go in the roughness of the alumina and form larger aggregates after sintering. An analogy can be made with the evaporation of emulsion droplets, presented by Manoharan et al.[91]. In their article, they studied a three-phase system where oil droplets containing small number (2 to 15) of microspheres at their surface dispersed in water form well-ordered clusters during drying due to the compressive force. Furthermore, since the same alumina concentration is employed but the Ludox nano-suspension leads to larger aggregates, it can be observed that the alumina walls are thinner in the latter suspension.

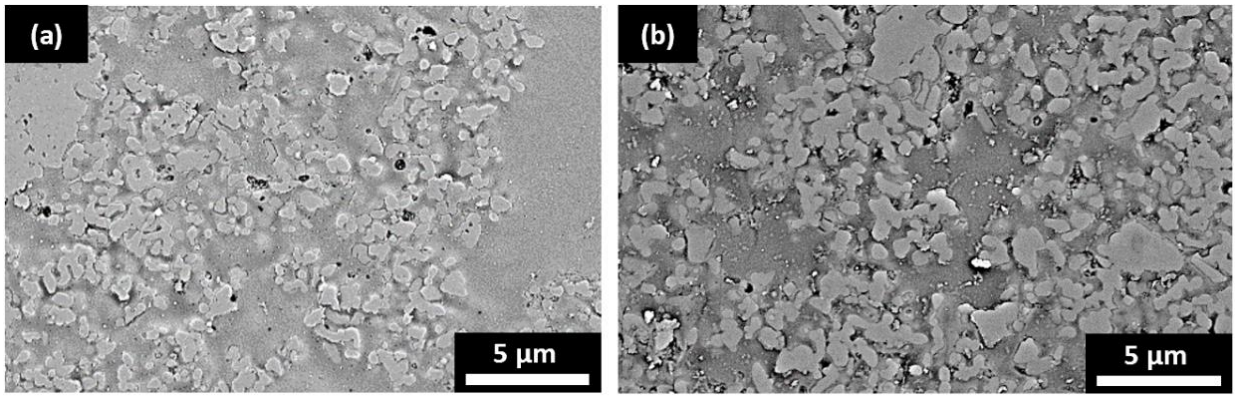

*Figure 8: SEM cuts of 10 vol% alumina bodies with (a) 0 vol% and (b) 2 vol% sucrose as a secondary fluid. There is the presence of pores (<5 μm) in between the alumina particles, confirming the existence of a hierarchical pore size.*

More precise SEM analysis was conducted to examine the microstructure of the alumina in greater detail. The results, obtained for 10 vol% alumina, depicted in Figure 8a (0% sucrose) and 8b (2% sucrose), confirmed the presence of an additional pore size in the micron-range within the ceramic body. This proves that our body possesses a hierarchical pore size, as initially desired. The larger pore size, resulting from camphene crystallisation, was found to range between 20 and 50 μm (obtained from image analysis). In contrast, the smaller pore size, originating from the particle network formed by capillary suspension, was less than 5 μm. As shown in Figure 8, no significant differences are observed in the microstructure between the two concentrations. In both images, the alumina grains exhibited partial connectivity due to network formed by the capillary suspension bringing the particles into contact and to their partial sintering.

### 3.3.2 β-TCP porous bodies microscopy

Following the examination of alumina, β-TCP was investigated due to its superior biocompatibility for bone tissue engineering. Therefore, similar experiments were conducted on the β-TCP system, however, a distinct behaviour is observed, consistent with the change from a pendular state suspension (for the alumina) to a capillary state suspension for the β-TCP.

In the SEM transverse cuts (Figure 9), no clearly defined particle network is present. Instead, small aggregates seem to be dispersed within the medium – camphene – lacking the formation of large pores. The size of the aggregates increases with increasing secondary fluid to 20-50 µm for the 1 vol% of the Ludox nano-suspension (Figure 9b) and aggregates as large as 200 µm are noticeable for 2 vol% (Figure 9c). While the formation of round and individual aggregates seems to be specific to the β-TCP powder, the size and dependence with the Ludox concentration appears to be consistent for both alumina and β-TCP: the Ludox nano-suspension induces an increase in powder aggregation. The reason is still not clear, but could be related to the higher concentration of silica nanoparticles (50 wt%, or approximately 30 vol%) further increasing after water evaporation during the heating treatment. The increase of aggregates with increasing amounts of Ludox is especially visible for the β-TCP system due to the nature of the capillary state characterised by a three-phase contact angle higher than 90°. In the capillary state, multiple particles surround the secondary liquid droplets and thereby form a network, as opposed to individual particles connected by concave bridges in the alumina system. As a result, the capillary state is more prone to aggregate formation during dispersion of the secondary liquid. This combined with the higher particle size for the β-TCP particles can explain the reason behind the enhanced effect of the Ludox on the β-TCP, compared to the alumina.

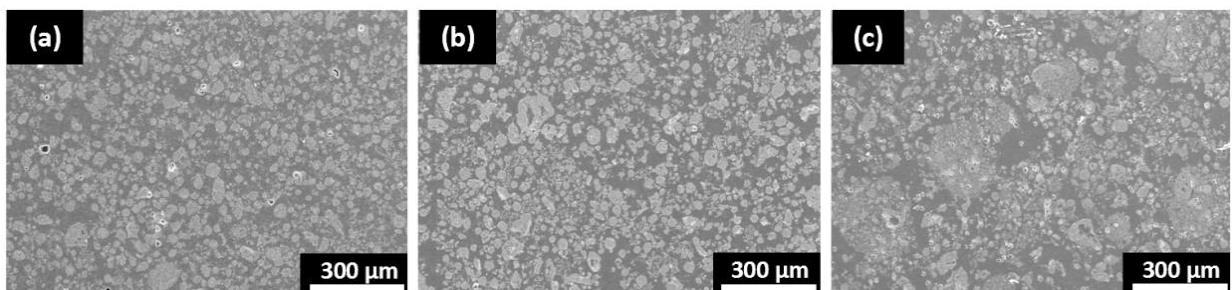

*Figure* 9: SEM images of 10 vol% *β-TCP* with (a) 0, (b) 1 and (c) 2 vol% of Ludox silica nano-suspension. The brighter colour represents the *β-TCP* while the darker one corresponds to the voids.

### 3.3.3 Pore characteristics of sintered samples

The porosities of our alumina and β-TCP scaffolds are calculated from the SEM images using a custom python code (present in the supplementary information), as shown in Figure 10. The error is calculated from the variance from multiple pictures with the exact same concentration. At first glance, it is evident that the scaffolds made with 7 vol% of alumina generally exhibit higher total porosity than those made with 10 vol%. This outcome was expected since the pore-making medium, i.e., camphene, occupies more space when the concentration of solid loading decreases. Consequently, it can develop more effectively before being sublimed, resulting in higher porosity when less powder is present in the capillary suspension. Comparing the different secondary fluids, however, shows that the porosity is influenced by more than just the solid loading.

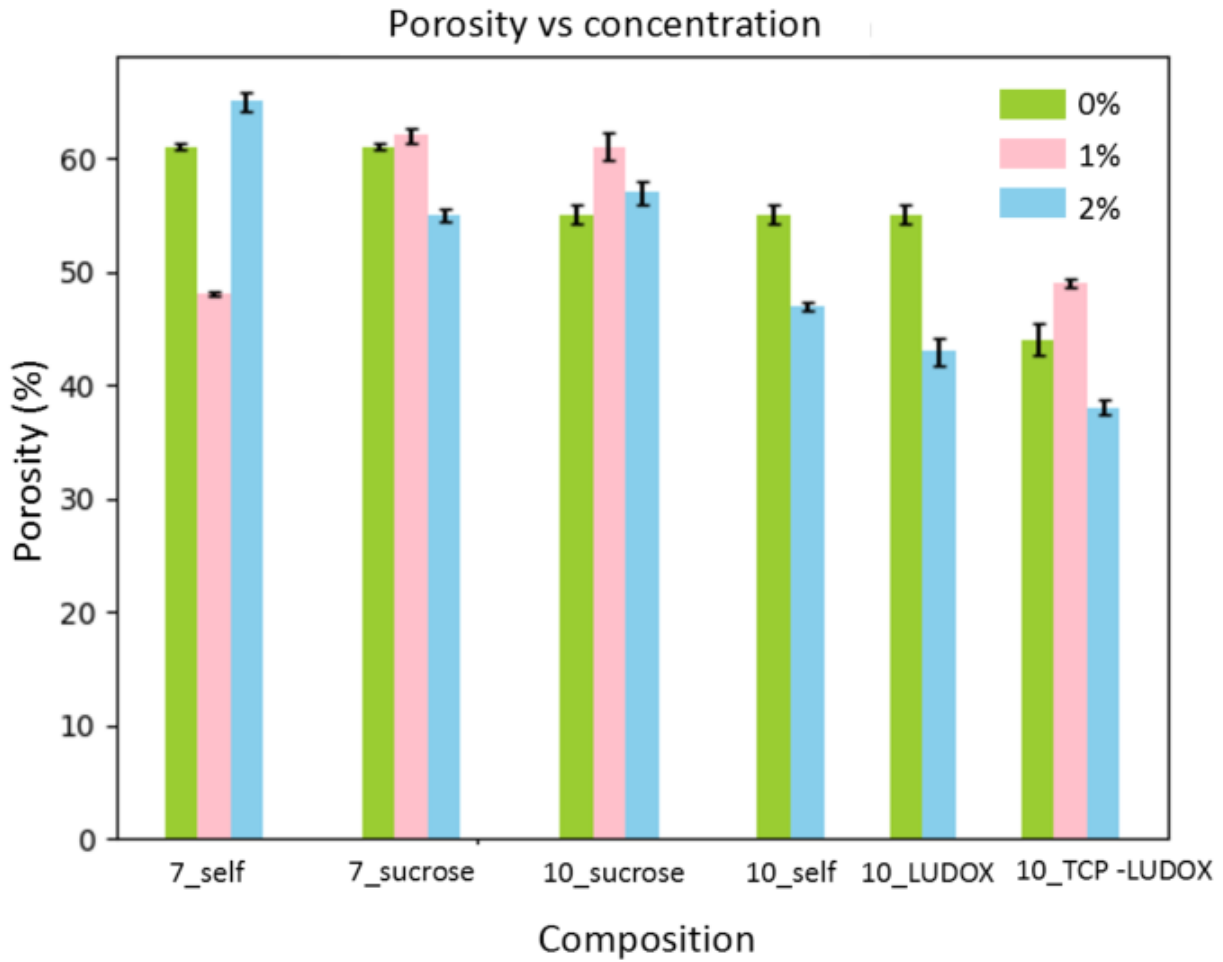

*Figure 10: Total porosity for different porous body concentrations including 7 vol% alumina with a self made silica nano-suspension, 7 vol% alumina with a sucrose solution, 10 vol% alumina with sucrose, 10 vol% alumina with a self made silica nano-suspension, 10 vol% alumina with a purchased Ludox silica nano-suspension, and β-TCP with the Ludox silica nano-suspension. The influence of the amount of secondary liquid are shown as different columns.*

For the 7 vol% alumina scaffolds, the porosities mostly lie in the 60% range, with the highest porosity observed for 2 vol% of the self-made silica nano-suspension and 1 vol% of sucrose as secondary fluids, with porosities of $65 \pm 0.8$ % and $62 \pm 0.6$ %, respectively. From the total porosity, it is again observed that the microstructure differs depending on whether the nano-suspension of silica or the sucrose is used. Previously, the SEM pictures for 7 vol% alumina with silica nano-suspension in Figure 5 revealed an overall similar microstructure, except when 2 vol% of secondary fluid was used, resulting in ellipsoidal pores. Comparing the porosities of such samples, we find that the porosity is the highest when 2 vol% is used ($65 \pm 0.8$ %) and is the lowest with 1 vol% ($48 \pm 0.2$ %), suggesting that having ellipsoidal pores is better for the total porosity than the branched ones. It confirms that there is an optimum concentration that provides both the highest total porosity with the desired microstructure.

On the contrary, when sucrose is used, the scaffold containing 1 vol% has a higher porosity ($62 \pm 0.6$ %) than the 2 vol% ($55 \pm 0.6$ %), even if the 2 vol% had ellipsoidal pores visible in the SEM images. We still do not know why this behaviour contrasts that of the silica nano-suspension.

For the 10 vol% alumina scaffolds, the porosities predominantly drop below 60% with 2 vol% of the self and purchased silica nano-suspensions even showing values below 50% ($43 \pm 1.2$ % for 2 vol% Ludox) to $61 \pm 1.2$ % for the 1 vol% sucrose. Comparing these values to the case where no secondary fluid was added ($55 \pm 0.8$ %), it is evident that the addition of silica-based nano-suspensions decreases the total porosity making the scaffold denser and less desirable for our study. Nonetheless, such low porosities can be useful for applications other than bone tissue engineering, such as selective filtration [44 92 93 94 95 96]. The reason behind the in porosity for 10 vol% and not for 7 vol% when the silica nano-suspensions are used can be due to the increase of aggregates, as previously discussed.

Interestingly, when 10 vol% alumina is combined with 1 vol% sucrose, the total porosity approaches that of the 7 vol% alumina porous bodies ($61 \pm 1.2$ % versus $65 \pm 0.8$ %). This observation confirms the impressions from the SEM images where Figure 6e seems to exhibit the highest amount of porosity among the 10 vol% samples. The reason for this still requires further investigation.

To further assess porosity, we utilised mercury infiltration porosimetry (MIP) in addition to image analysis on 10 vol% alumina samples with 0 and 1 vol% sucrose. MIP results showed porosities of $60 \pm 0.5$% and $66 \pm 2$% for the 0 vol% and 1 vol% sucrose samples, respectively, which are slightly higher than those obtained from image analysis, likely because MIP accounts for wall porosity, which is more challenging to capture with image analysis.

Apart from the total porosity, the distribution of pore sizes and their anisotropy is important for bone implant applications. These values could not be calculated from the β-TCP samples as they did not have the right microstructure and, therefore, only the alumina samples are characterised here. The equivalent pore size was obtained using the diameter of 2D circular pores of the same area. The Feret diameter, which corresponds to the longest distance between two extremities, gives an indication of the pore elongation.

The pore size distribution shows that more than 94% of the pores have a diameter between 10 and 150 μm with the remaining pores being between 200 and 1000 μm. This small pore size range is shown in Figure 11, while the Feret diameter ratio and the entire pore size histograms can be found in Supplemental Figures S11 and S12, respectively. The first row of histograms represents the pore size distribution (diameter) for the concentrations of 7 vol% alumina with increasing amount of the self-made silica nano-suspension (0, 0.5, 1, 2 and 5 vol%). The second row shows the histograms of two different solid loadings: the two first graphs stand for 7 vol% alumina with 1 and 2 vol% of sucrose. The last and third row displays pore size distributions calculated from image analysis for a solid loading concentration of 10 vol% with respectively 0, 1 and 2 vol% of sucrose, followed by 2 vol% of Ludox and self-made silica nano-suspensions.

Upon examination of the different histograms, it can be observed that all the distributions follow the same trend: they present a peak for the lower pore sizes (~ 20 µm) and then an exponential-like decrease towards the higher pore sizes. The pore size obtained for all the concentrations seem to have a similar pore size distribution, even if different concentrations were tested. The different capillary suspensions should give different particle network since the capillary force would also be different, resulting in pores of different size. However, the magnitude of the capillary force given by different secondary liquid does not seem to be sufficient to result in a modification made by the pores coming from the growing dendrites.

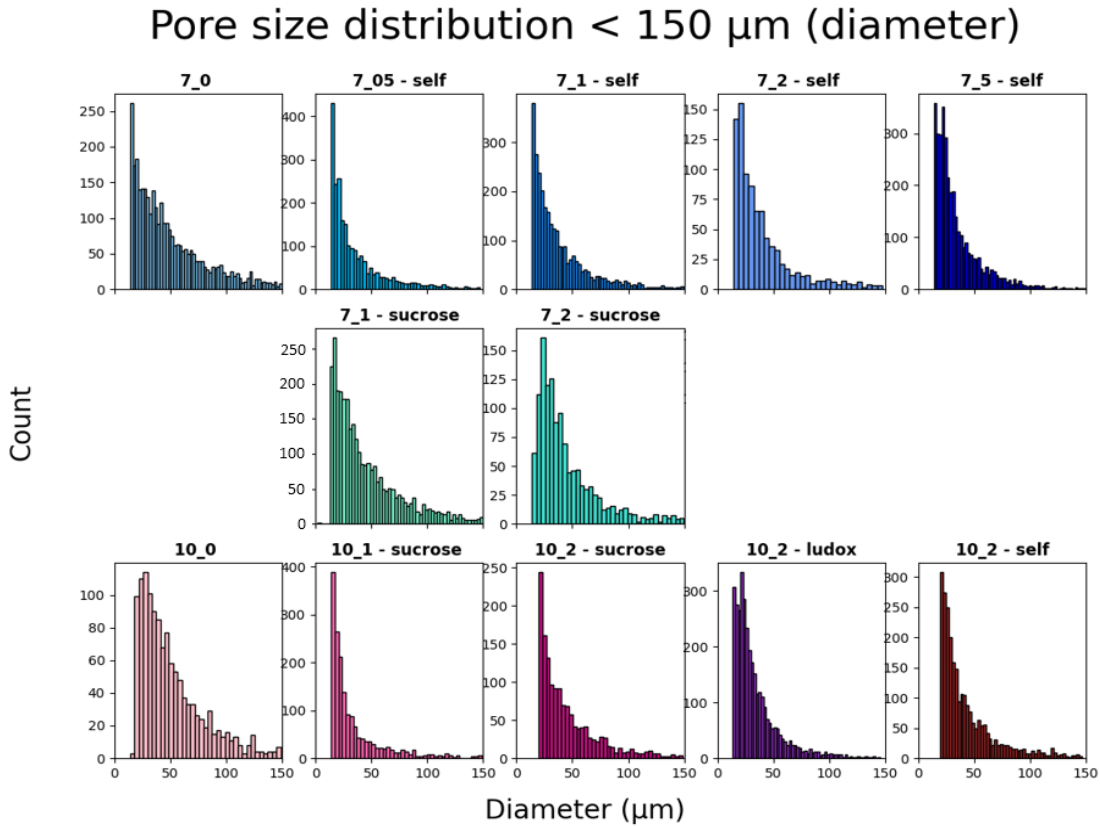

*Figure 11: Pore size distribution histogram (magnified below 150 µm) for alumina materials. The title of each histogram represents the concentration of the material: the first number stands for the solid loading concentration, the second number and the last word account for the concentration and the type of secondary fluid used, respectively.*

Since the thermal gradient used during freeze casting was the same for all the concentrations, the resulting pore size distribution is the same. Nonetheless, there is a slight shift in the pore size peak, e.g. between the 7 vol% alumina with sucrose as a secondary liquid, which may be indicative of the capillary network. Since the network is expected to primarily influence the small pores (as shown in Figure 8), this change may not be captured in the detected pore size distribution.

The logarithmic differential intrusion, representing the volume of mercury introduced into the sample as a function of pore size obtained from mercury infiltration porosimetry (MIP), is shown in Supplementary Figure S13 for samples containing 10 vol% alumina with 0 and 1 vol% sucrose. Unlike image analysis, two peaks are observed: one in the range of 0.1-1 µm and another at 10-50 µm. The small pore sizes likely correspond to wall pores undetectable by image analysis, as these sizes match those seen in high magnification images of the alumina wall (Figure 8). The larger pore sizes correlate with those obtained from image analysis, representing pores formed by camphene sublimation. For some samples, a third peak is observed between 0.7 and 3 µm in the MIP pore size distribution. Since the entire samples are characterized using MIP, the denser core also influences the distribution. We believe that peak emerges because of the intermediate pore size in the scaffolds centre, region that is neglected in the image analysis. This indicates that MIP is a more precise method for characterizing pores as it can detect both types. However, MIP could only be conducted for two concentrations and only for alumina samples since $\beta$-TCP is too brittle for MIP and partially dissolves in water and ethanol.

The pore size analysis from both techniques (image analysis and MIP) give different but complementary information about the scaffolds' microstructures. Furthermore, the feasibility and coherence of MIP experiments reaffirm the presence of open and interconnected pores in the material.

The application targeted in this work is the creation of bone grafts for bone tissue engineering, with targeted pore sizes of 100-500 µm [24,7,97,98,99,100,101] for bone vascularisation and osseogenesis, as well as 20-50 µm for osteoblasts proliferation and migration [97,102,101]. Reaching a high porosity (> 60%) and pore interconnectivity are additionally of utmost importance for fluid transport and viability of the grafts [5,103,100]. While our scaffolds exhibit high porosities, interconnectivity, and multimodal pore size distributions, the latter fail to entirely align with the intended pore sizes. Notably, the largest pore size range achieved is 25-50 µm, suitable for osteoblast proliferation, whereas the smallest pore range falls between approximately 0.5 to 5 µm, measured from inter-grain distance (Figure 7). This size is inadequate for bone tissue engineering but remains applicable in cell culture [104,105,106] and filtration [107,108,92] applications. Research by Nittami et al. [107] suggests that polytetrafluoroethylene (PTFE) membranes with 0.5 and 1 µm pores are effective in wastewater treatment, exhibiting reduced fouling. Additionally, Kim et al. [104] demonstrated the efficacy of 0.9 µm membranes for cell culture, boasting superior characteristics over commercially available options, including enhanced cell imaging quality and increased cell-cell contact

Moreover, the pore size of our samples can be modified by altering the cooling rate of camphene during solidification, thereby manipulating the thermal gradient. To investigate this, experiments were conducted, wherein camphene was crystallised under different freezing conditions. The outcomes of these experiments are presented in the Supplemental Figure S4. Lower temperatures result in reduced spacing between individual dendrites, leading to smaller pores, as is also confirmed in freeze casting literature [55,56]. Conversely, there is a higher degree of interconnection between dendrites at a higher temperature of 40°C, thereby enhancing pore interconnectivity. These experiments highlight that the final pore characteristics within the porous body can be altered by adjusting the freezing temperature.

The distribution of the ratio between the Feret diameter $d_{Feret}$ and the equivalent circular pore diameter $d_{area}$ is shown in Supplemental Figure S11. All the concentrations present a ratio between 1 and 3, indicating that mostly elongated pores are present. This confirms the structures obtained from the SEM pictures, where both dendritic and ellipsoidal pores can be considered as elongated. That ratio was quantified into three categories: $\frac{d_{Feret}}{d_{area}} \leq 1.5$, $1.5 < \frac{d_{Feret}}{d_{area}} < 2.5$ and $\frac{d_{Feret}}{d_{area}} \geq 2.5$. These separations were chosen to distinguish between circular pores with a diameter ratio less than 1.5 and elongated with a diameter ratio greater than 2.5. The results are presented in Table 2. They indicate that the scaffolds made of 7 vol% with 2% and 0.5% self-made silica nano-suspension and all the 10 vol% scaffolds possess the lowest percentage of circular pores, and subsequently the highest percentage of elongated pores. Their higher percentage in elongated pores implies a higher Feret diameter compared to the equivalent circular pore. The SEM pictures show a lower degree of interconnectivity (because of a lack of secondary dendrites) for those structures. Pores coming from dendritic structures are wider than the ellipsoidal ones, leading to a lower ratio between the Feret diameter and the equivalent pore size.

| Composition | $\frac{d_{Feret}}{d_{area}} \leq 1.5$ (%) | $1.5 < \frac{d_{Feret}}{d_{area}} < 2.5$ (%) | $\frac{d_{Feret}}{d_{area}} \geq 2.5$ (%) |
|---|---|---|---|
| 7% alumina | 62.9 | 36.6 | 0.5 |
| 7% alumina + 0.5% self | 32.5 | 65.3 | 2.2 |
| 7% alumina + 1% self | 48.8 | 50.1 | 1.1 |
| 7% alumina + 2% self | 32.0 | 65.3 | 2.7 |
| 7% alumina + 5% self | 45.6 | 53.2 | 1.2 |
| 7% alumina + 1% sucrose | 59.1 | 40.4 | 0.5 |
| 7% alumina + 2% sucrose | 36.1 | 61.4 | 2.5 |
| 10 % alumina | 31.9 | 66.3 | 1.8 |
| 10% alumina + 1% sucrose | 31.6 | 65.7 | 2.7 |
| 10% alumina + 2% sucrose | 28.5 | 69.5 | 2.0 |
| 10% alumina + 2% LUDOX | 23.2 | 70.7 | 6.1 |
| 10% alumina + 1% self | 25.5 | 69.4 | 5.1 |

*Table 2: Ratio of the Feret diameter $d_{Feret}$ to the equivalent circular pore size $d_{area}$, as split into three domains.*

The Feret angle calculations (Figure S14) confirm this hypothesis: almost all the distributions of the Feret angle present a peak at 45°. This corresponds to the way the secondary dendrites form when camphene crystallises. Indeed, the secondary dendrites form at an angle of 45° from the primary dendrite. This indicates that the image analysis can successfully describe the reality. The only concentrations that do not present this peak at 45°, but possess a flatter Feret angle distribution, are the same concentrations as the ones that have the higher ratio between the Feret diameter and the equivalent pore size: the 7 vol% with 2 vol% sucrose and the 10 vol%. This confirms that those concentrations do not tend to form secondary dendrite but have an elongated shape.

# 4 Conclusion

In this work, for the first time, freeze casting of a capillary suspension is applied successfully as a novel method to realise hierarchical porous structures. To this end, three different secondary liquids are examined: sucrose, a self-made and a commercial silica nano-suspension, with two kinds of ceramics, inside a camphene environment. The alumina system is found to lie in the pendular state while the β-TCP seems to be in the capillary state, explaining the difference in their microstructure after sintering.

Both the amount and the kind of secondary liquid affect the microstructure and porosity. However, only the type of secondary liquid used affects the rheology: when used with alumina, sucrose leads to a lower moduli as opposed to its effects on β-TCP. Our study highlights that for 7 vol% of alumina and 2 vol% of any secondary liquid, a transition from dendritic to ellipsoidal shaped pores appears. This corresponds to a strengthening of the spanning particle network of the capillary suspension.

All the freeze casting experiments are performed at room temperature, but it is possible to tune the pore size distribution by changing the thermal gradient. The finally obtained porous structures offer both a high level of porosity and pore interconnectivity. Our scaffolds possess a fully open porosity displaying two main pore size distributions. In a wide variety of application such as filtration, energy storage and (bone) tissue engineering, it is essential to have both. Our research results in porous scaffolds with 25-50 μm pores connected channels. Moreover, it was proven that the walls exhibit pores with size <5 μm, useful for cell culture and filtration.

More investigations need to be done, especially for the β-TCP system, where more secondary liquids can be tested, and their microstructure identified. Another way to further investigate the β-TCP system would be to decrease the particle size by ball-milling and to find other secondary liquids that permit to obtain a pendular state. Finally, better mixing procedures aiming to enhance capillary suspension homogenisation and avoid air bubbles should be studied. Before the proposed porous bodies can be used as implants, they must undergo extensive testing to evaluate their biological response. First, the biocompatibility of the materials and associated immune system response should be evaluated. Second, the vascularization and osteogenesis should be assessed.

# 5   Acknowledgments

The authors would like to thank financial support from Odysseus Program (grant agreement no. G0H9518N), the International Fine Particle Institute (IFPRI) and European Union's Horizon 2020 research and innovation programme Marie Skłodowska-Curie grant agreement No 955612.

We also would like to thank Dr. Ir. Anja Vananroye and Carlo Everaerts for their technical support in making the freeze casting set-up, Lingyue Liu for aiding with the confocal images and Mia Kovač for helping in making β-TCP platelets, useful for three-phase contact angle measurements.

# Supplemental Information

for

# Hierarchical materials with interconnected pores from capillary suspensions for bone tissue engineering

Souhaila Nider* [1], Femke De Ceulaer [1], Berfu Goksel [2], Annabel Braem [2], Erin Koos* [1]

[1] KU Leuven, Department of Chemical Engineering, Leuven, Belgium

[2] KU Leuven, Department of Materials Engineering, Leuven, Belgium

* Corresponding authors: erin.koos@kuleuven.be, souhaila.nider@kuleuven.be

## Table of Contents

# 1. Supplemental figures

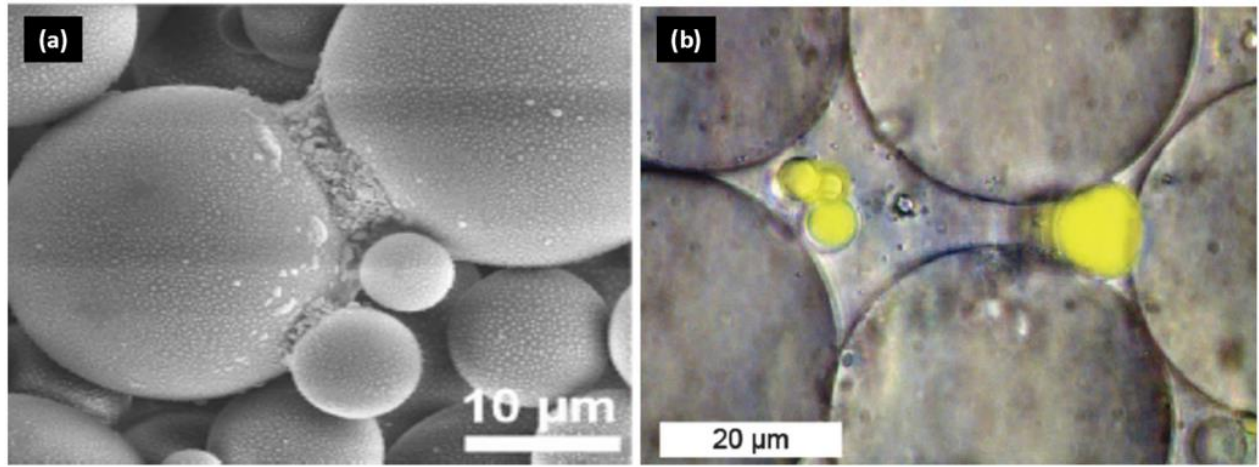

Figure S1: (a) Pendular state obtained from Hauf et al. [83] with concave secondary liquid bridges connecting individual particles (b) Capillary state obtained from Koos et al. [82] with convex secondary liquid connecting multiple particles.

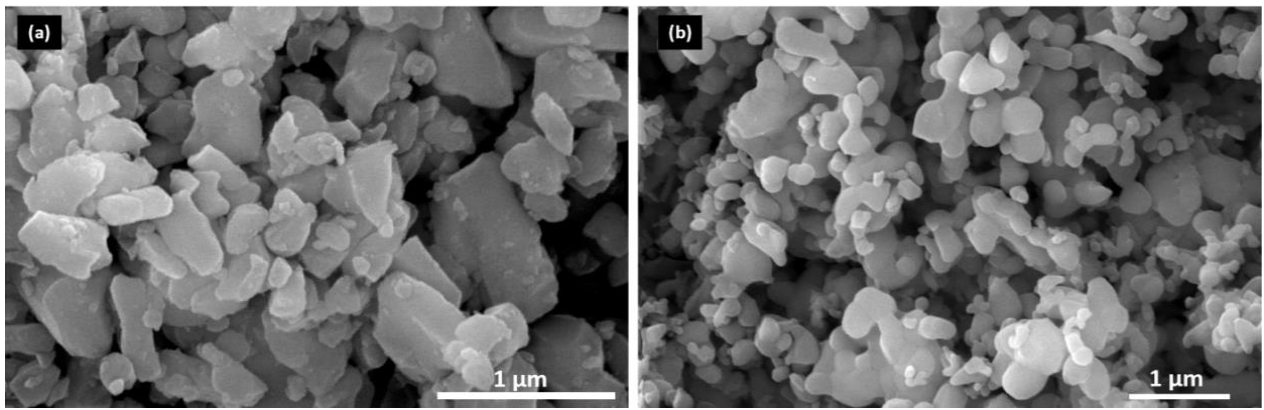

Figure S2: Raw materials used in our study (a) alumina powder (b) porous β-TCP granules.

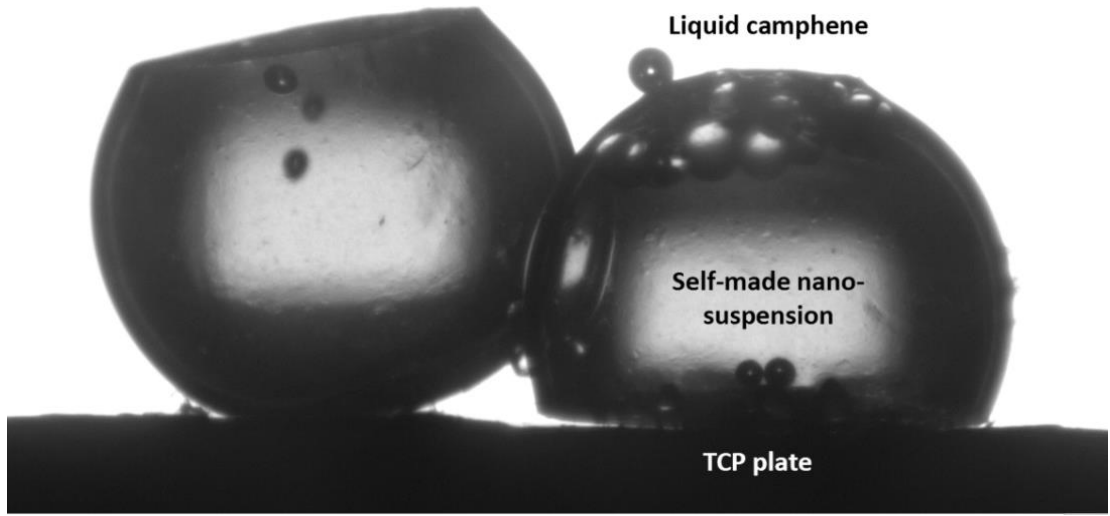

Figure S3: Self-made silica nano-suspension on a β-TCP plate in liquid camphene after agitation. The droplets maintained their original shape due to the gelled particle network inside each sessile drop.

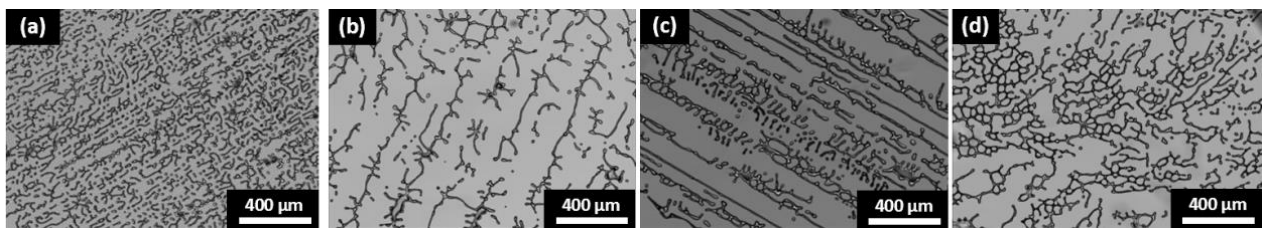

Figure S4: Study of camphene crystallisation at (a) -12°C, (b) 8°C, (c) 20°C and (d) 40°C.

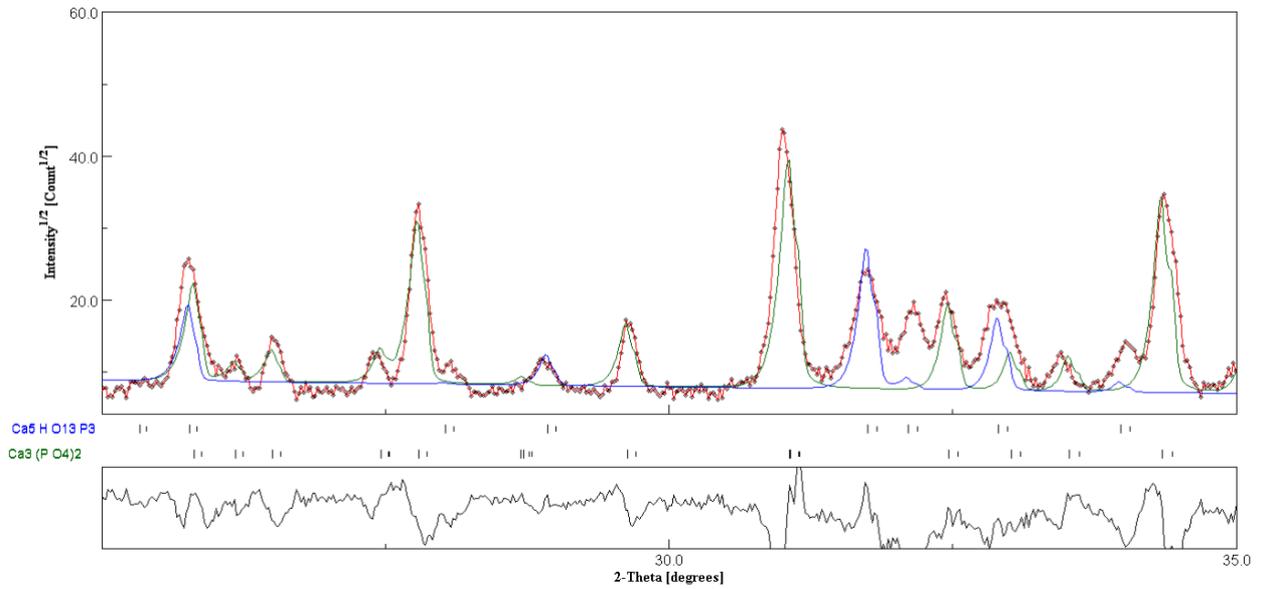

Figure S5: XRD spectrum of the purchased β-TCP powder without any modification. The raw data are in black. The beta phase (green curve) is present, along with another calcium phosphate phase.

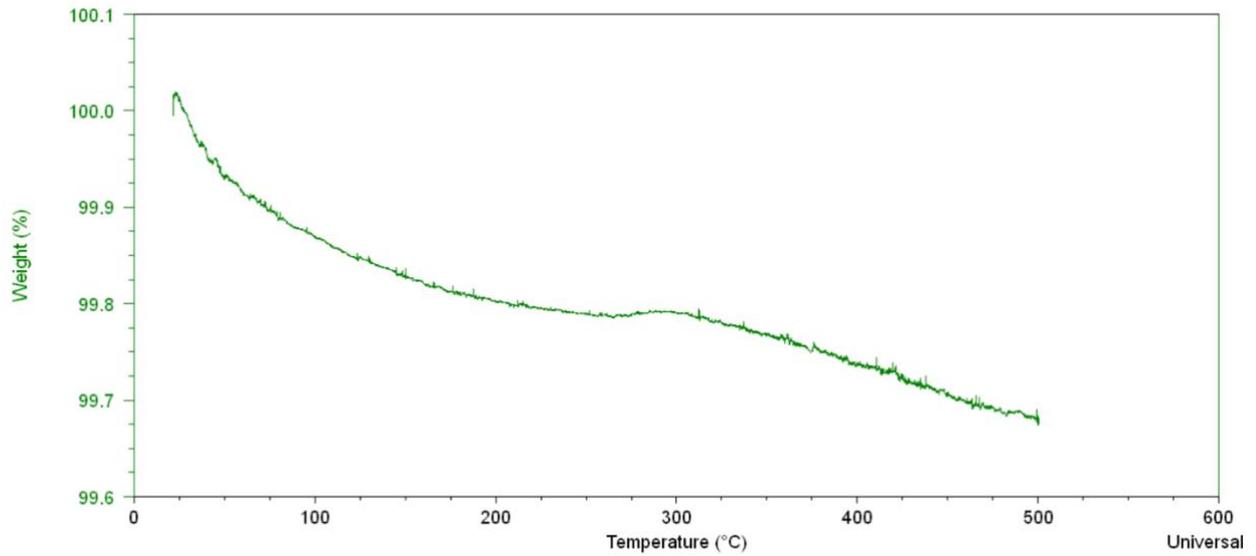

Figure S6: Thermogravimetric analysis on β-TCP powder in the 20-500°C temperature range, conducted with a heating rate of 10°C/min. The decrease in the weight % is insignificant, indicating that no organic coating was used on the powder.

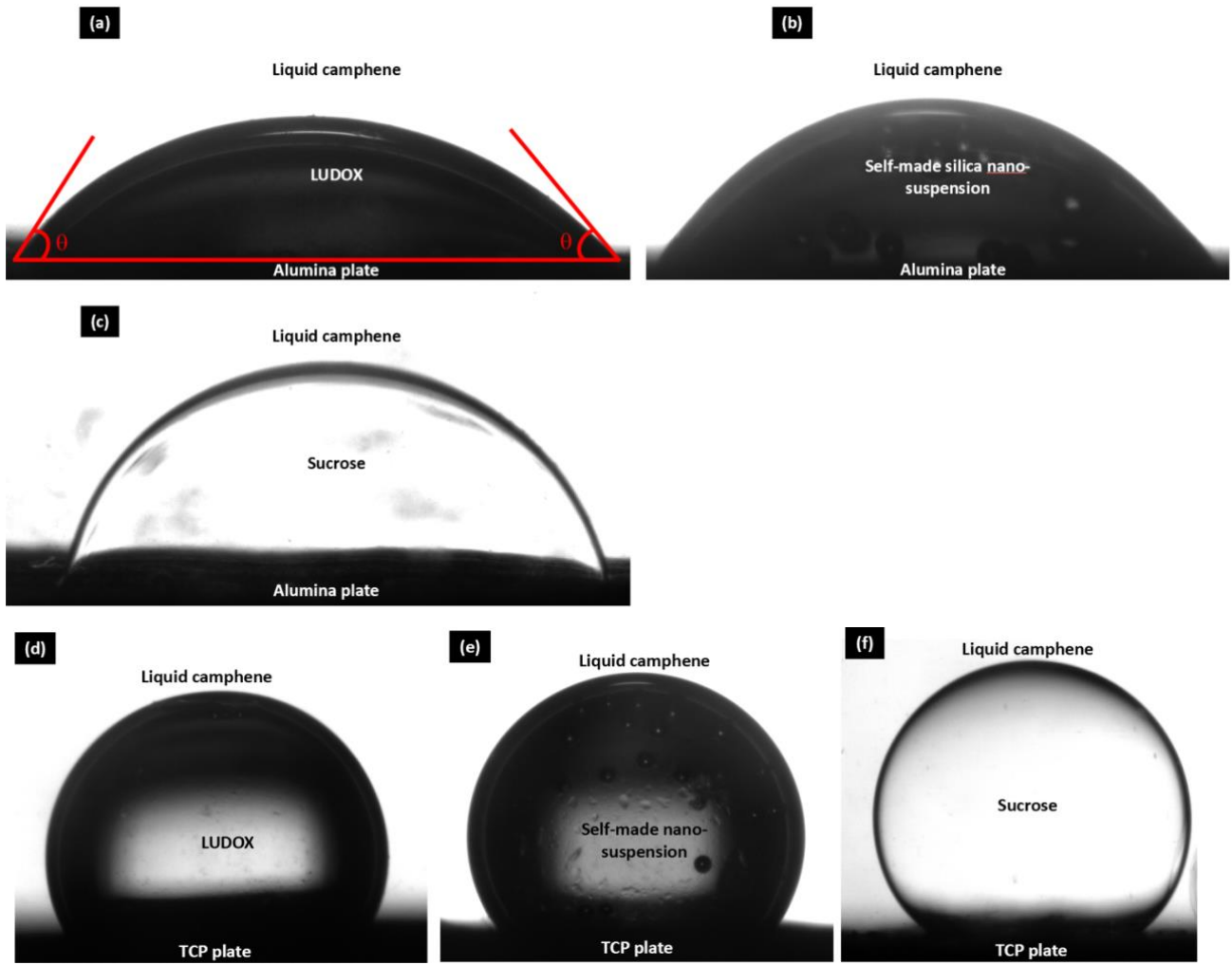

Figure S7: Three-phase contact angles of (a) a LUDOX silica nano-suspension, (b) a self-made silica nano-suspension, (c) a sucrose solution on an alumina plate inside a camphene environment, (d) a LUDOX silica nano-suspension, (e) a self-made silica nano-suspension, (f) a sucrose solution on a β-TCP plate inside a camphene environment. To ensure the liquid state of camphene, the experiment was conducted at 60°C. The sucrose presents a higher contact angle than the silica nano-suspensions. An example of the contact line is highlighted in red in picture (a), where $\theta$ indicates the three-phase contact angle.

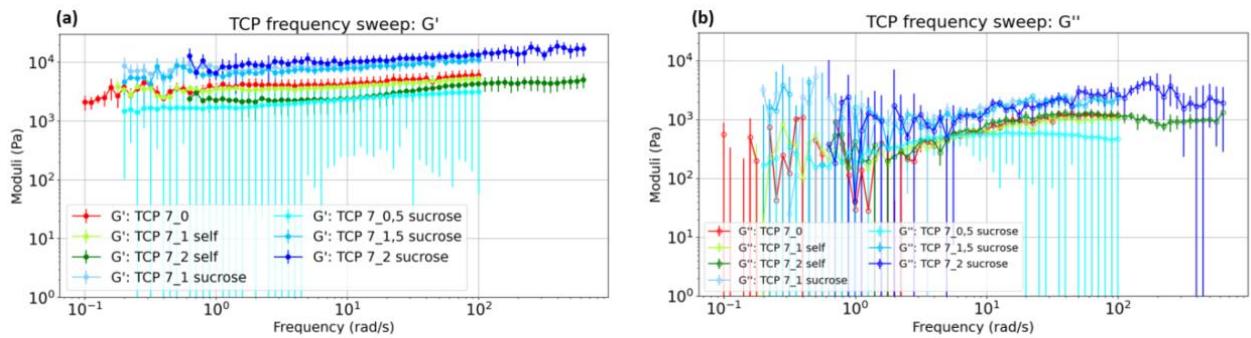

Figure S8: β-TCP frequency sweep showing (a) $G'$ and (b) $G''$ as a function of applied frequency. The notation "self" stands for "self-made silica nano-suspension".

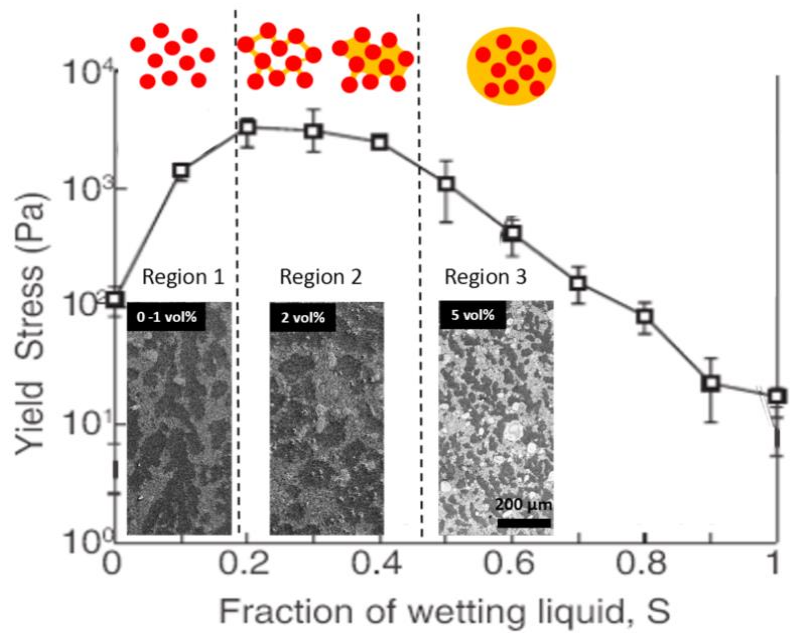

Figure S9: Adapted figure from Koos and Willenbacher [Science, 2011, 331(6019), 897-900], yield stress evolution of a hematite in DINP/water system depending on the wetting liquid fraction S. This graph shows that with an increase amount of secondary liquid, the yield stress increases (spanning particle network formation) until reaching a plateau (the arrangement of particles stays the same) before dropping down (formation of aggregates not linked to each other). Our corresponding microstructure are added to the graph.

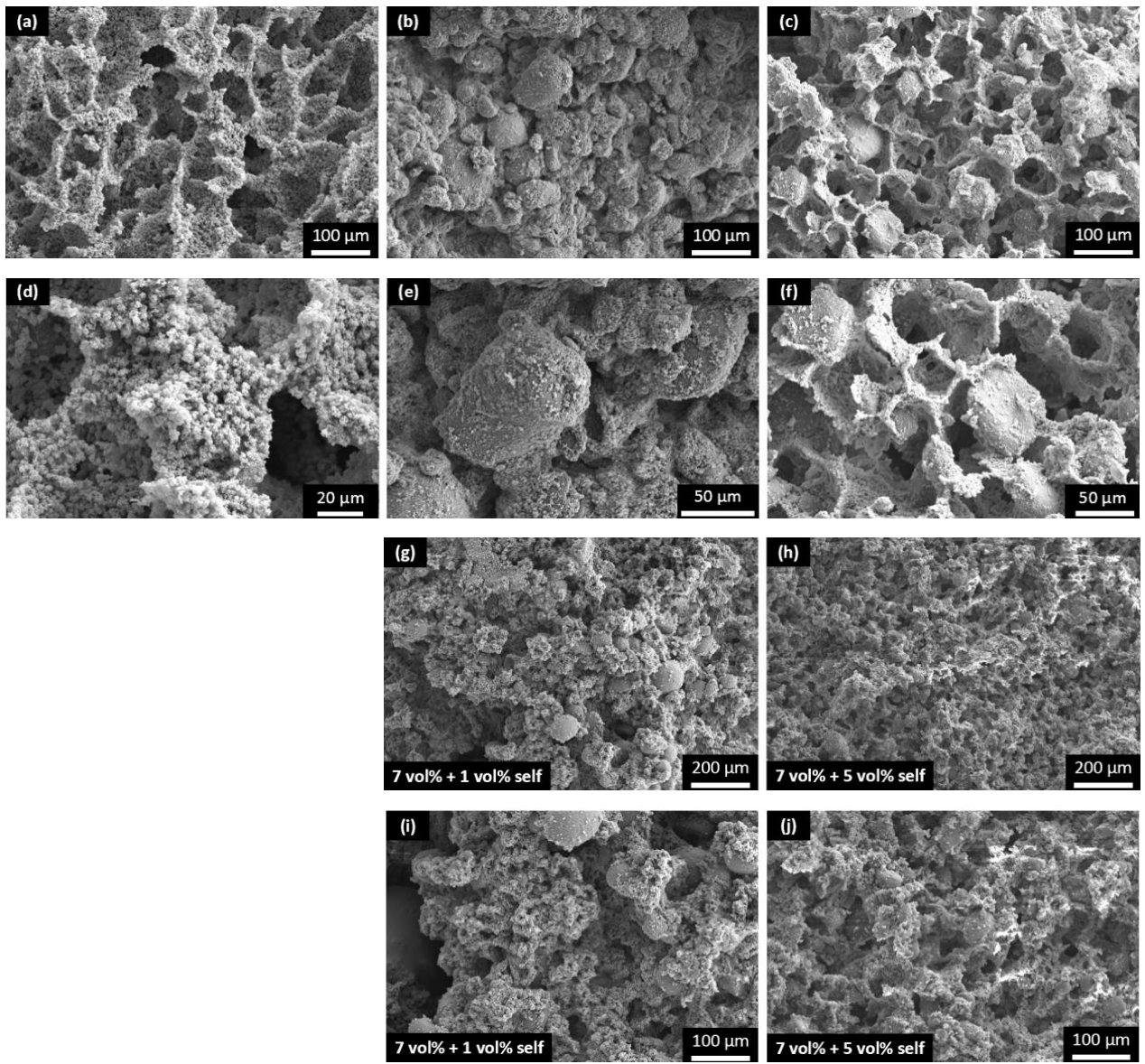

Figure S10: SEM images of fractured samples composed of 7 vol% alumina with (a)(d): a 0 vol%, (b)(e): 1 vol% sucrose, (c)(f): 2 vol% sucrose, (g)(i): 1 vol% self-made silica nano-suspension, (h)(j): 2 vol% self-made silica nano-suspension.

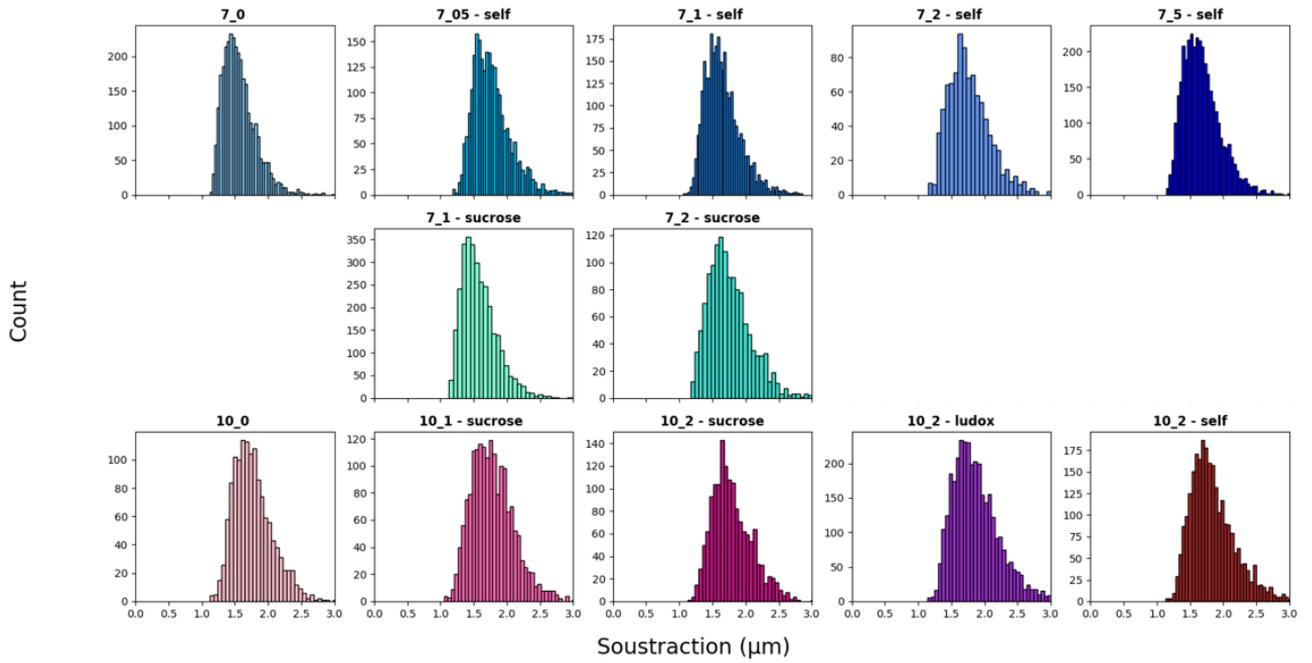

Figure S11: Histogram of the ratio of the Feret diameter with the equivalent circular pore size.

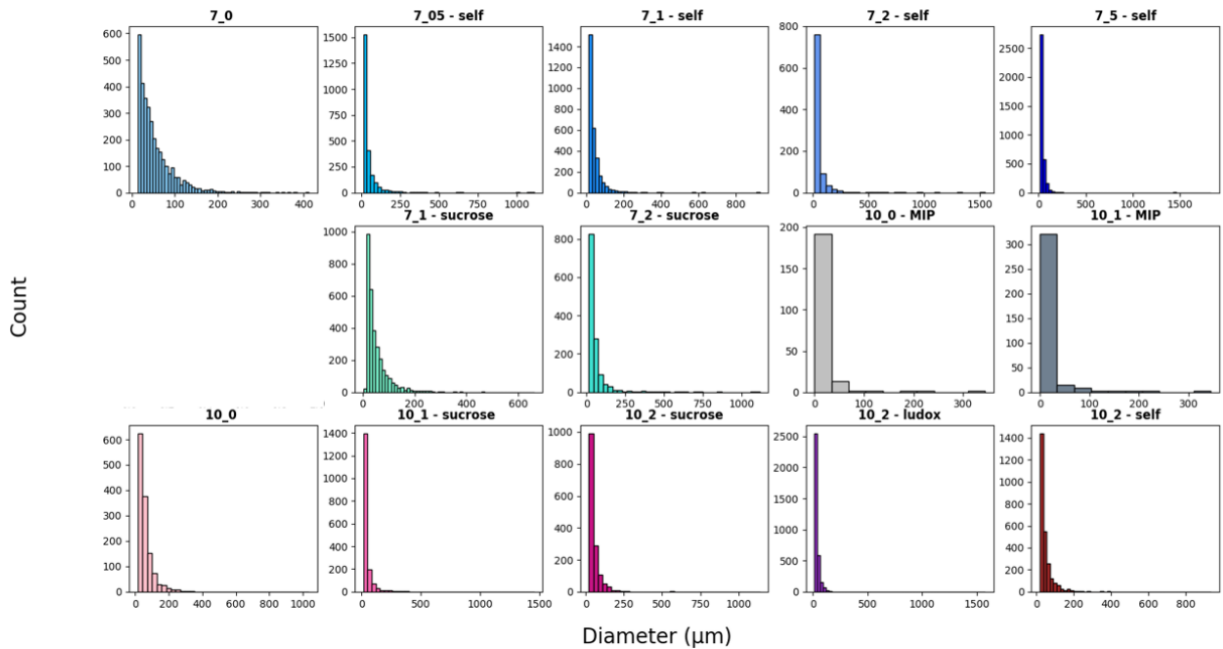

Figure S12: Histogram of the entire pore size distribution.

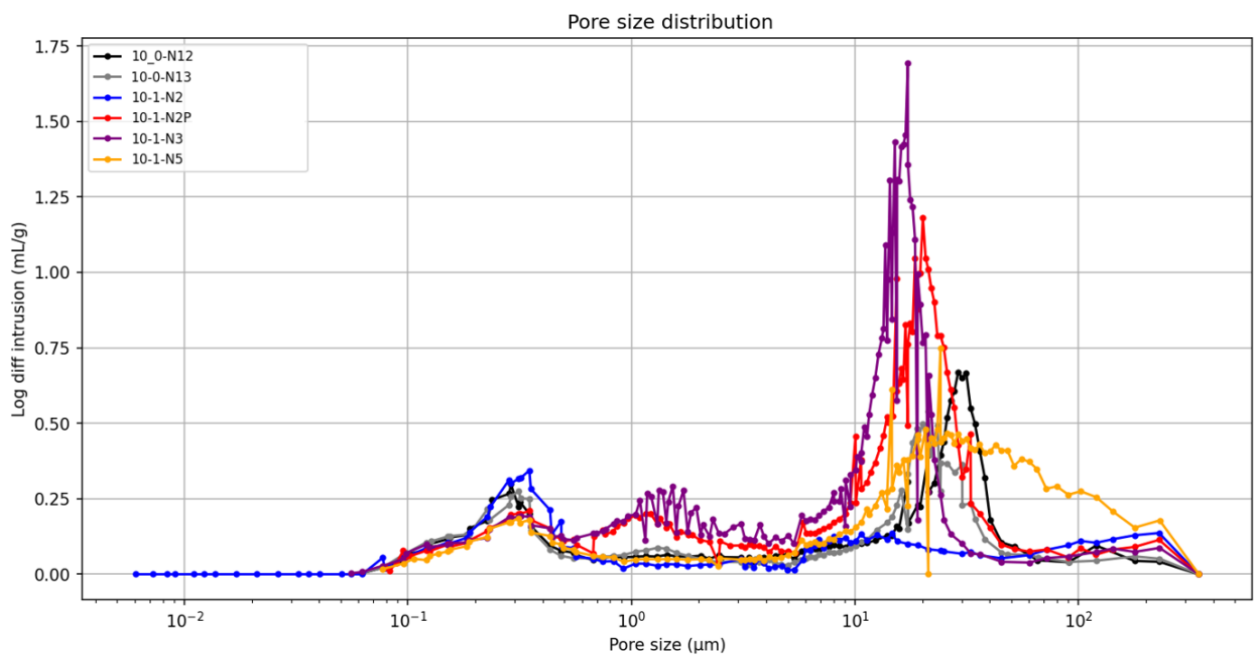

Figure S13: Logarithmic differential intrusion from mercury infiltration porosimetry for samples containing 10 vol% alumina with 0 (10_0) and 1 vol% (10_1) sucrose.

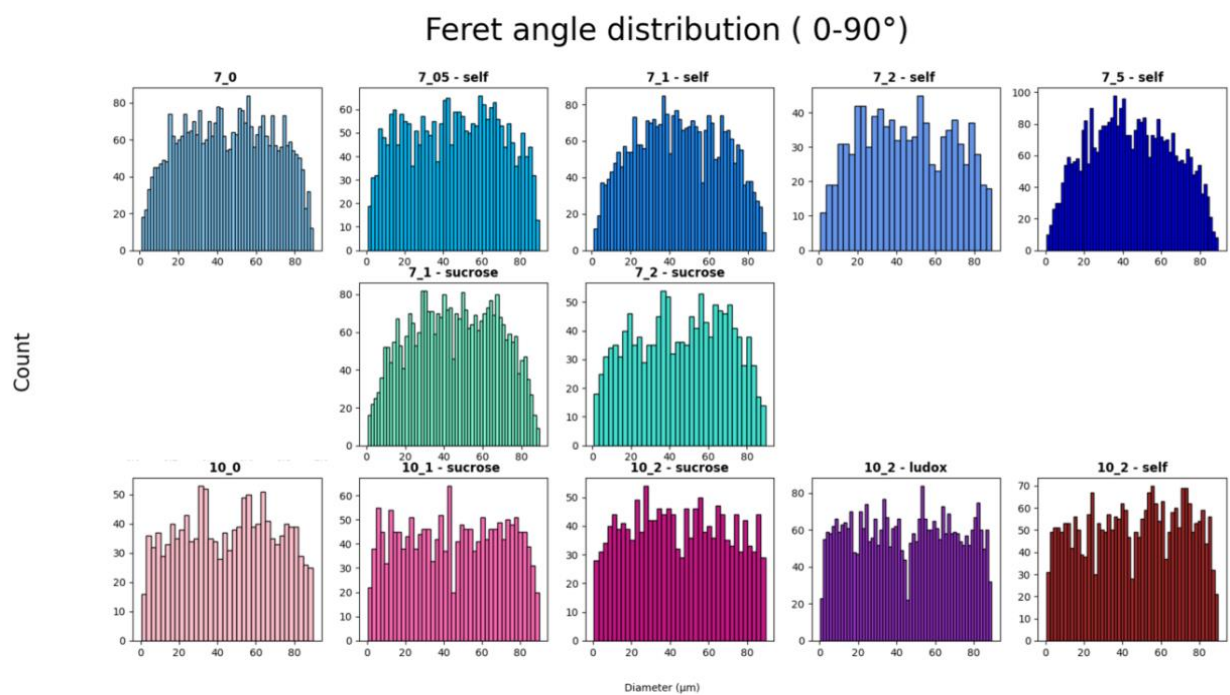

Figure S14: Feret angle distribution.

## 2. Results from data fitting using β-TCP (Ca$_3$(PO$_4$)$_2$) and Hydroxyapatite (Ca$_5$(PO$_4$)$_3$(OH))

Red line – Overall fitting

Blue line – Hydroxyapatite

Green line – β-TCP

Full scan:

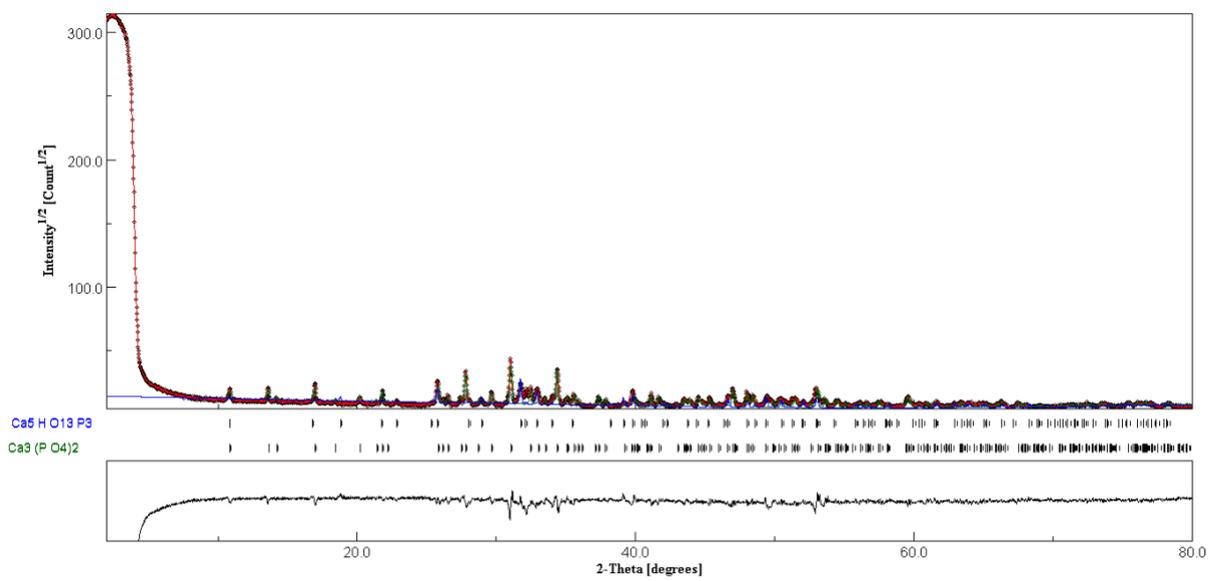

Scan range: 10°-60°

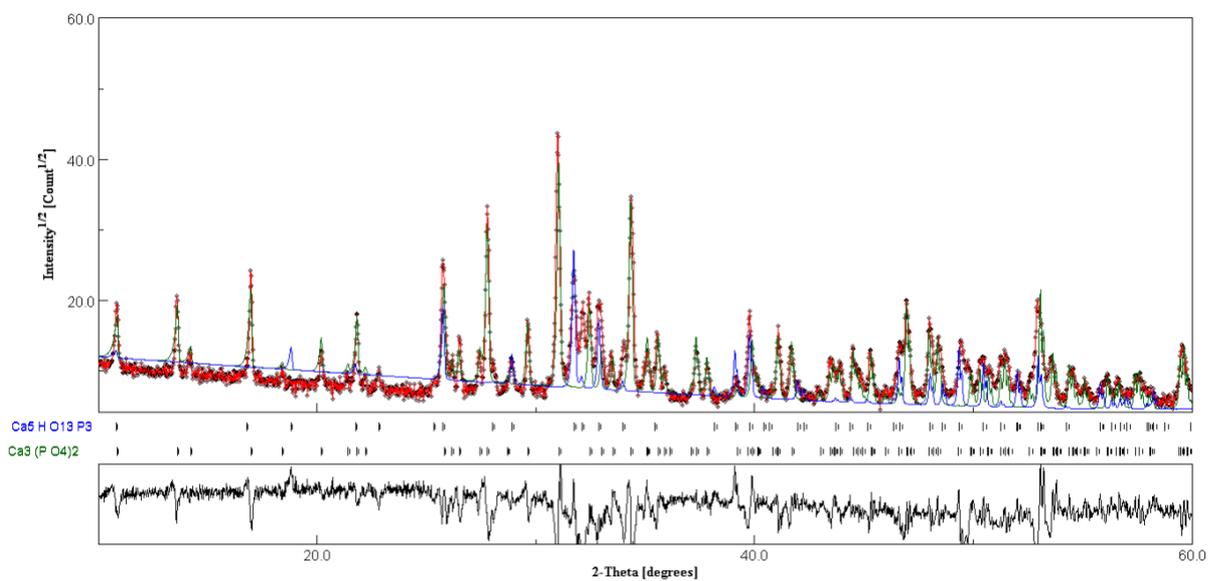

## Scan range: 25°-35°

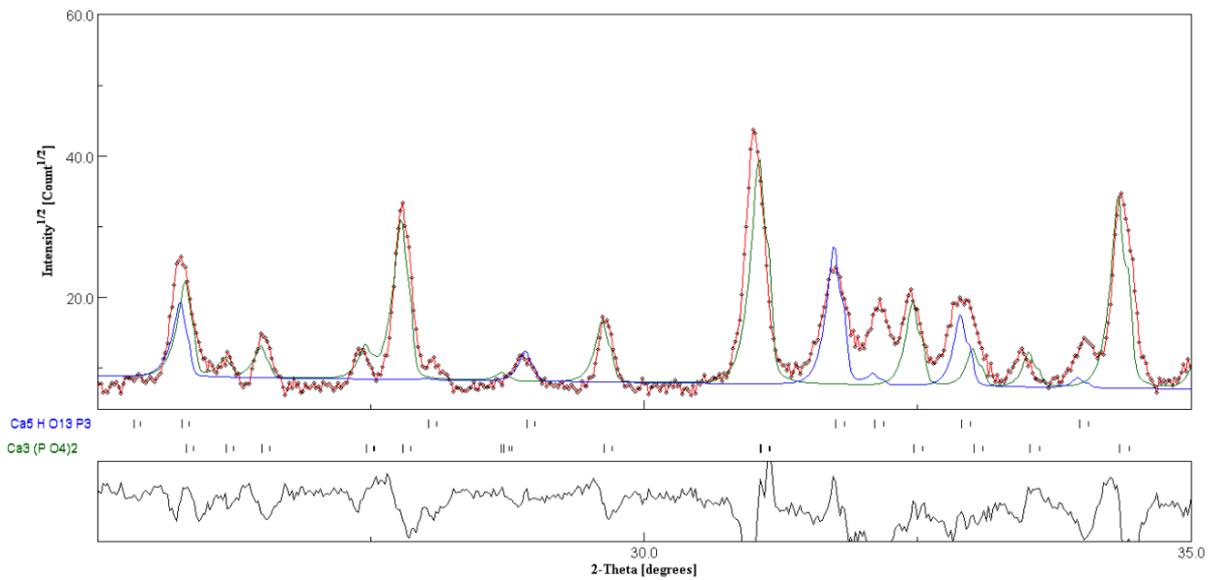

```
5, iterations - pause
sig= 47.867813
Rwp (%) = 97.99187
Rwpnb (%, no bkg) = 99.52421
Rwpnb1 (%, no bkg rescaled) = 99.26956
Rwpnb2 (%, no bkg rescaled^2) = 99.07997
Rb (%) = 96.985825
Rexp (%) = 2.0471349
# iterations = 5
Ca3 (P O4)2 , weight %: 4.9324417 +- 0.0
 Ca5 H O13 P3 , weight %: 95.06756 +- 47.429127
```

```
Full Pattern Search Match
Tube radiation: Cu
Final radiation: 1.5406
Using database: COD_all.sqlite
Elements to include: O Ca P
Structures to include: Mineral Inorganic Organic Metalorganic Unknown
Start filtering from database...
Only phases with density > 0.95 are accepted!
Cell parameter refinement > isotropic expansion/contraction allowed!
Number of structures filtered from database: 130
Number of structures to test: 130
Filtering time: 1 s
Weighting scheme: -1
Weight for Rietveld: 0.333333
Weight for smooth fit: 0
Weight for 1st deriv: 0.333333
Weight for 2nd deriv: 0.333333
Using Durbin-Watson for range setting.
Control parameter: -1.4
Structures # in this cycle: 130
Phase found: 7(O4 P), 11(Ca) 30351.2
Cycle finished: Actual wss: 3.04334e+06, number of phases :1
Phases found: 1529466, 7(O4 P), 11(Ca), vol.: 1, cryst: 1016.42, : 5.71412e-05
Structures # in this cycle: 130
Phase found: Calcium Tris(phosphate(V)) Hydroxide 312496
Cycle finished: Actual wss: 3.04143e+06, number of phases :2
Phases found: 1529466, 7(O4 P), 11(Ca), vol.: 0.729584, cryst: 1006.98, : 4.85953e-05
Phases found: 2300273, Calcium Tris(phosphate(V)) Hydroxide, vol.: 0.270416, cryst: 999.905, : 0.000444877
Structures # in this cycle: 94
Phase found: Ca, 2(O3 P1) 464090
Cycle finished: Actual wss: 3.0898e+06, number of phases :3
Phases found: 1529466, 7(O4 P), 11(Ca), vol.: 0.808471, cryst: 1005.29, : 2.41951e-06
Phases found: 2300273, Calcium Tris(phosphate(V)) Hydroxide, vol.: 0.166086, cryst: 1001.18, : 0.00017858
Phases found: 4000621, Ca, 2(O3 P1), vol.: 0.0254425, cryst: 999.475, : 0.000176423
Structures # in this cycle: 53
Phase found: O2 1.11927e+06
Cycle finished: Actual wss: 3.12165e+06, number of phases :4
Phases found: 1529466, 7(O4 P), 11(Ca), vol.: 0.64047, cryst: 1003.34, : 0.00039878
Phases found: 2300273, Calcium Tris(phosphate(V)) Hydroxide, vol.: 0.211131, cryst: 1000.14, : 0.000377939
Phases found: 4000621, Ca, 2(O3 P1), vol.: 0.0243826, cryst: 1000, : 0.000371784
Phases found: 1528336, O2, vol.: 0.124017, cryst: 999.907, : 0.000645753
Structures # in this cycle: 44
Phase found: Ca9.2 (P O4)5.55 F2 1.56783e+06
Cycle finished: Actual wss: 3.12123e+06, number of phases :5
Phases found: 1529466, 7(O4 P), 11(Ca), vol.: 0.61334, cryst: 1006.59, : 0.000272599
Phases found: 2300273, Calcium Tris(phosphate(V)) Hydroxide, vol.: 0.184667, cryst: 999.49, : 0.000293367
Phases found: 4000621, Ca, 2(O3 P1), vol.: 0.0117066, cryst: 999.981, : 0.00038366
Phases found: 1528336, O2, vol.: 0.102692, cryst: 999.932, : 0.000554825
Phases found: 2300587, Ca9.2 (P O4)5.55 F2, vol.: 0.0875941, cryst: 1000.03, : 0.000610501
Structures # in this cycle: 27
Phase found: Ca4(PO4)2O 1.84237e+06
Cycle finished: Actual wss: 3.04113e+06, number of phases :6
Phases found: 1529466, 7(O4 P), 11(Ca), vol.: 0.648861, cryst: 1018.2, : 0.000246416
Phases found: 2300273, Calcium Tris(phosphate(V)) Hydroxide, vol.: 0.187457, cryst: 1001.07, : 0.000212455
Phases found: 4000621, Ca, 2(O3 P1), vol.: 0.00814669, cryst: 1000.4, : 0.0649074
Phases found: 1528336, O2, vol.: 0.0751979, cryst: 999.865, : 0.00094286
Phases found: 2300587, Ca9.2 (P O4)5.55 F2, vol.: 0.0676404, cryst: 999.952, : 0.000891731
Phases found: 9011144, Ca4(PO4)2O, vol.: 0.0126975, cryst: 999.704, : 0.000457131
Structures # in this cycle: 24
Phase found: O 1.78193e+06
Cycle finished: Actual wss: 3.04249e+06, number of phases :6
Phases found: 1529466, 7(O4 P), 11(Ca), vol.: 0.531774, cryst: 1004.3, : 1.39691e-05
Phases found: 2300273, Calcium Tris(phosphate(V)) Hydroxide, vol.: 0.171026, cryst: 999.663, : 0.000303422
Phases found: 9011144, Ca4(PO4)2O, vol.: 0.0604439, cryst: 999.977, : 0.000393322
Phases found: 1528336, O2, vol.: 0.0729315, cryst: 999.873, : 0.0000638087
Phases found: 2300587, Ca9.2 (P O4)5.55 F2, vol.: 0.0744102, cryst: 999.863, : 0.000457413
Phases found: 1512527, O, vol.: 0.0894149, cryst: 1000.17, : 0.00115013
Structures # in this cycle: 23
Phase found: Ca3(PO4)2 2.26324e+06
Cycle finished: Actual wss: 3.04157e+06, number of phases :7
Phases found: 1529466, 7(O4 P), 11(Ca), vol.: 0.578828, cryst: 1009.98, : 7.10469e-05
Phases found: 2300273, Calcium Tris(phosphate(V)) Hydroxide, vol.: 0.175666, cryst: 1001.37, : 0.000290633
Phases found: 9011144, Ca4(PO4)2O, vol.: 0.00787444, cryst: 999.878, : 0.000344678
Phases found: 1528336, O2, vol.: 0.0566462, cryst: 999.914, : 0.000906112
Phases found: 2300587, Ca9.2 (P O4)5.55 F2, vol.: 0.0545746, cryst: 1000.67, : 0.000771668
Phases found: 1512527, O, vol.: 0.0787603, cryst: 1000.44, : 0.000310426
Phases found: 9005865, Ca3(PO4)2, vol.: 0.0476512, cryst: 999.76, : 0.00127207
```

Final wss: 3.04157e+06
Time to search and quantify: 21 s

Found phases and quantification:

| Phase COD ID | cif file | name | vol. (%) | wt. (%) | crystallites (Å) | microstrain |
|---|---|---|---|---|---|---|
| 1529466 | 1529466.cif | 7(O4 P), 11(Ca) | 57.8828 | 64.2849 | 1009.98 | 7.10469e-05 |
| 2300273 | 2300273.cif | Calcium Tris(phosphate(V)) Hydroxide | 17.5666 | 17.5995 | 1001.37 | 0.000290633 |
| 9011144 | 9011144.cif | Ca4(PO4)2O | 0.787444 | 0.784008 | 999.878 | 0.000344678 |
| 1528336 | 1528336.cif | O2 | 5.66462 | 3.08799 | 999.914 | 0.000906112 |
| 2300587 | 2300587.cif | Ca9.2 (P O4)5.55 F2 | 5.45746 | 4.47049 | 1000.67 | 0.000771668 |
| 1512527 | 1512527.cif | O | 7.87603 | 3.88614 | 1000.44 | 0.000310426 |
| 9005865 | 9005865.cif | Ca3(PO4)2 | 4.76512 | 5.88699 | 999.76 | 0.00127207 |

Final Rietveld analysis, Rw: 0.971267, GofF: 28.1219

## Final Rietveld fit

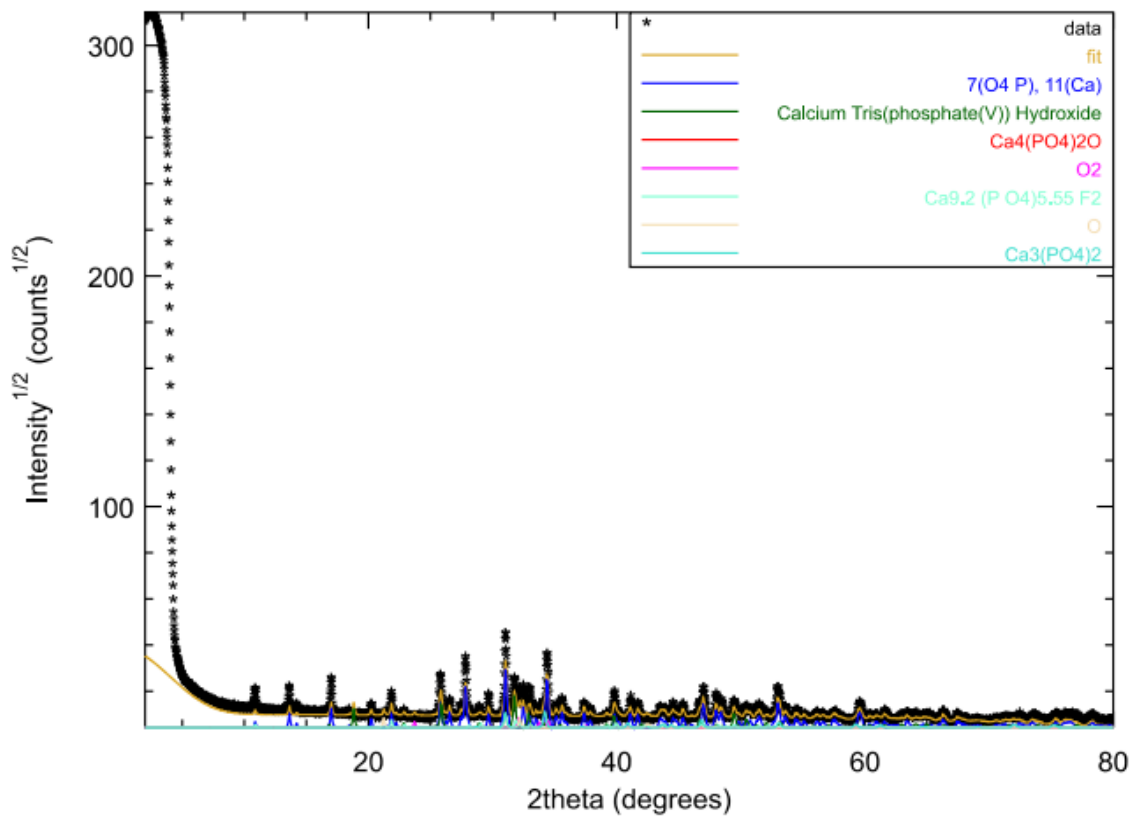

FPSM computation finished!

# 3. Python code for the porosity calculation

```python
#Code that will only apply a bilateral blur and an Otsu filter and calculates the resulting porosity
import cv2
import os
import shutil
import numpy as npimport statistics
#from tiffile import *
import matplotlib.pyplot as plt
#List to store porosities
porosity1_list = []
porosity2_list = []
#Loop through all pictures in input directory
input_dir = "File to analyse - 10_1 sucrose"
output_dir = "Analysed pictures - 10_1 sucrose"
for filename in os.listdir(input_dir):
    if filename.endswith(".jpg") or filename.endswith(".tif") or filename.endswith(".png"):
        img = cv2.imread(os.path.join(input_dir, filename))
        height, width= img.shape[:2]
        #img = img[:heihgt - 60,  :] # if I want to crop the picture
        # Application of bilateral filter
        blur_bilateral =  cv2.bilateralFilter(img, 13, 45, 45)
        # Application of median blur filter to make a background of intensity
        #background = cv2.medianBlur(blur_bilateral, 301)
        # We need to substract the backroun from the bilateral filtered image so that the gradient in intensity goes away
        #difference = cv2.subtract(blur_bilateral, cv2.add(background, -20))
        # Convert the image to grayscale
        #gray_img = cv2.cvtColor(img, cv2.COLOR_BGR2GRAY)
        blur_bilateral = cv2.cvtColor(blur_bilateral, cv2.COLOR_BGR2GRAY)
        #difference = cv2.cvtColor(difference, cv2.COLOR_BGR2GRAY)
        #Otsu thresholding on the original image
        #ret1, th1 = cv2.threshold(gray_img, 0, 255, cv2.THRESH_BINARY+cv2.THRESH_OTSU)
        #Otsu thresholding on the bilateral filtered image
        ret2, th2 = cv2.threshold(blur_bilateral, 0, 255, cv2.THRESH_BINARY+cv2.THRESH_OTSU)
        #Otsu threholding on the gaussian filtering (meaning the substracted picture)
        #ret3, th3 = cv2.threshold(difference, 0, 255, cv2.THRESH_BINARY+cv2.THRESH_OTSU)
        # Cleaning/ de-noising of the resulting Otsu filers by delating and then eroding
        #cleaned1 = cv2.dilate(cv2.erode(th3, (7,7)), (7,7))
        #cleaned2 = cv2.bitwise_not(cv2.dilate(cv2.erode(cv2.bitwise_not(cleaned1), (7,7)), (7,7)))
        cleaned1 = cv2.dilate(cv2.erode(th2, (100,100)), (100,100))
        cleaned2 = cv2.bitwise_not(cv2.dilate(cv2.erode(cv2.bitwise_not(cleaned1), (100,100)), (100,100)))
        #porosity calculation
        white_pixels1 = np.sum(cleaned1 == 255)
        black_pixels1 = np.sum(cleaned1 == 0)
        white_pixels2 = np.sum(cleaned2 == 255)
        black_pixels2 = np.sum(cleaned2 == 0)
```

```python
        porosity1 = black_pixels1 / (black_pixels1 + white_pixels1)
        porosity2 = black_pixels2 / (black_pixels2 + white_pixels2)
        porosity1_list.append(porosity1)
        porosity2_list.append(porosity2)
        #save pictures
        #cv2.imwrite(os.path.join(output_dir, f"{filename[:-4]}_img.tif"), img)
        #cv2.imwrite(os.path.join(output_dir, f"{filename[:-4]}_blur_bilateral.tif"), blur_bilateral)
        #cv2.imwrite(os.path.join(output_dir, f"{filename[:-4]}_background.tif"), background)
        #cv2.imwrite(os.path.join(output_dir, f"{filename[:-4]}_difference.tif"), difference)
        #cv2.imwrite(os.path.join(output_dir, f"{filename[:-4]}_th3.tif"), th3)
        cv2.imwrite(os.path.join(output_dir, f"{filename[:-4]}_cleaned1_100.tif"), cleaned1)
        cv2.imwrite(os.path.join(output_dir, f"{filename[:-4]}_cleaned2_100.tif"), cleaned2)
        cv2.imwrite(os.path.join(output_dir, f"{filename[:-4]}_th2.tif"), th2)
    #save porosities in txt file
with open(os.path.join(output_dir, "Only bilateral and Otsu-total_porosities1_et2.txt"), 'a') as f:
    print(porosity1_list, statistics.variance(porosity1_list))
    f.write("{}    {:.2f}%     +-    {:.4f}\n   ".format(filename,    sum(porosity1_list)/len(porosity1_list),
statistics.variance(porosity1_list)))
    print(porosity2_list, statistics.variance(porosity2_list))
    f.write("{}    {:.2f}%     +-    {:.4f}\n   ".format(filename,    sum(porosity2_list)/len(porosity2_list),
statistics.variance(porosity2_list)))

def total_porosity_calculation(input_dir):
    porosity_list=[]
    for filename in os.listdir(input_dir):
        if  filename.endswith(".tif") or filename.endswith(".PNG"):
            img = cv2.imread(os.path.join(input_dir, filename))
            black_pixels = np.sum(img == 0)
            white_pixels = np.sum(img == 255)
            porosity = black_pixels/(black_pixels + white_pixels)
            porosity_list.append(porosity)
            #with open(os.path.join(input_dir, "total_porosities_cropped.txt"), 'a') as f:
                #f.write("{} {:.2f}% \n".format(filename, porosity))
    with open(os.path.join("result cropped porosities", "total_porosities_cropped_10_sucrose.txt"), 'a') as f:
        f.write("{}    {:.2f}%     +-    {:.4f}\n   ".format(str(input_dir),    sum(porosity_list)/len(porosity_list),
statistics.variance(porosity_list)))
'''
total_porosity_calculation("Analysed pictures - 7_0 cropped")total_porosity_calculation("Analysed pictures
- 7_1 self cropped")total_porosity_calculation("Analysed pictures - 7_1 sucrose
cropped")total_porosity_calculation("Analysed pictures - 7_2 self
cropped")total_porosity_calculation("Analysed pictures - 7_2 sucrose
cropped")total_porosity_calculation("Analysed pictures - 7_05 self
cropped")total_porosity_calculation("Analysed pictures - 7_5 self
cropped")total_porosity_calculation("Analysed pictures - 10_0
cropped")total_porosity_calculation("Analysed pictures - 10_2 ludox
cropped")total_porosity_calculation("Analysed pictures - 10_2 self
```

cropped")total_porosity_calculation("Analysed pictures - 10_1 sucrose cropped")total_porosity_calculation("Analysed pictures - 10_2 sucrose cropped")

## 4. Reasons for selecting the current mixing technique for our capillary suspension preparation

Capillary suspension preparation methods were extensively explored by Bossler et al. (F. Bossler, L. Weyrauch, R. Schmidt, and E. Koos. "Influence of mixing conditions on the rheological properties and structure of capillary suspensions." *Colloids and Surfaces A: Physicochemical and Engineering Aspects* 518 (2017): 85-97), which we used as a reference to test various mixing procedures adapted to our system. After significant optimisation, we selected the following procedure: melting camphene (bulk fluid) and the secondary fluid together at 60°C, then mixing them at high frequencies using an ultrasonic mixer. Subsequently, pre-heated (60°C) powder is added, and all components are mixed at 700 rpm with a propeller mixer for 3 hours at 60°C.

Other methods for preparing capillary suspensions were ineffective in achieving a homogenised suspension in our case. For instance, mixing the powder with the bulk fluid using a speed mixer was not feasible for us because camphene must stay above its solidification temperature of 48°C. We also attempted mixing the particles with camphene in a water bath at 700 rpm before adding the pre-heated secondary liquid dropwise, but this resulted in large secondary liquid spheres (~5 mm diameter) due to insufficient droplet breakup, higher affinity of the secondary fluid with alumina and its immiscibility with the bulk fluid.

Using the ultrasonic mixer on the pure suspension before adding the secondary liquid or mixing all components together at once also led to visible agglomerates. Our current method of ultrasonic mixing of both fluids before adding the powder and subsequent propeller mixing for three hours produced a visibly homogenised suspension without aggregates. Shorter mixing times resulted in sedimentation.

While there is room for improvement, this preparation method was the best given the constraints, as other techniques like speed mixing and planetary ball-milling were not possible due to the need to maintain camphene at 60°C throughout the process.

All the preparation methods were tested with the alumina system. Since we found a technique that seemed to effectively work, we applied the same procedure to the TCP system.